\documentclass[aps,floatfix,showpacs,preprintnumbers,amsmath,amssymb]{revtex4}
\usepackage{natbib}
\usepackage{dcolumn}
\usepackage{graphicx}
\usepackage{enumerate}
\usepackage{hyperref}

\newcommand{\mnras}{Mon.~Not.~R.~Astron.~Soc.}

\newcommand{\phrd}{Phys.~Rev.~D.} 
\newcommand{\jcap}{J.~Cosmol.~Astropart.~Phys.} 
\newcommand{\be}{\begin{equation}}
\newcommand{\ee}{\end{equation}}
\newcommand{\bea}{\begin{eqnarray}}
\newcommand{\eea}{\end{eqnarray}}
\def\[{\begin{equation}}
\def\]{\end{equation}}
\begin{document}
\title{Figures of merit and constraints from testing General Relativity using the latest cosmological data sets including refined COSMOS 3D weak lensing}
\author{Jason N.  Dossett\footnote{Electronic address: jnd041000@utdallas.edu}, Jacob Moldenhauer\footnote{Electronic address: jam042100@utdallas.edu}, Mustapha Ishak\footnote{Electronic address: mishak@utdallas.edu}}
\affiliation{
Department of Physics, The University of Texas at Dallas, Richardson, TX 75083, USA;}
\date{\today}
\begin{abstract}
{We use cosmological constraints from current data sets and a figure of merit approach in order to probe any deviations from general relativity at cosmological scales. The figure of merit approach is used to study and compare the constraining power of various combinations of data sets on the modified gravity (MG) parameters. We use the recently refined HST-COSMOS weak-lensing tomography data, the ISW-galaxy cross correlations from 2MASS and SDSS luminous red galaxy surveys, the matter power spectrum from SDSS-DR7 (MPK), the WMAP7 temperature and polarization spectra, the baryon acoustic oscillations from Two-Degree Field and SDSS-DR7, and the Union2 compilation of type Ia supernovae, in addition to other bounds from Hubble parameter measurements and big bang nucleosynthesis. We use three parametrizations of MG parameters that enter the perturbed field equations. In order to allow for variations of the parameters with the redshift and scale, the first two parametrizations use recently suggested functional forms while the third is based on binning methods. Using the first parametrization, we find that the CMB + ISW + WL combination provides the strongest constraints on the MG parameters followed by CMB + WL or CMB + MPK + ISW. Using the second parametrization or the binning methods, we find that the combination CMB + MPK + ISW consistently provides some of the strongest constraints. This shows that the constraints are parametrization dependent. We find that adding up current data sets does not improve consistently the uncertainties on MG parameters due to tensions between the best-fit MG parameters preferred by different data sets. Furthermore, some functional forms imposed by the parametrizations can lead to an exacerbation of these tensions. Next, unlike some studies that used the CFHTLS lensing data, we do not find any deviation from general relativity using the refined HST-COSMOS data, confirming previous claims in those studies that their result may have been due to some systematic effect. Finally, for all the parametrizations and binning methods used, we find that the values corresponding to general relativity are within the $95\%$ confidence level contours for all data set combinations.} 
\end{abstract}
\maketitle
\section{introduction}
One of the routes to understanding the origin of cosmic acceleration is by testing whether the acceleration is due to a dark energy component pervading the universe or some modification to general relativity (GR) at cosmological scales of distances. The study of the growth rate of large scale structure can help to make such a distinction and has been the subject of a number of analyses in the literature. Some first papers (see for example \cite{ScoccimarroStarkman2004,Song2005a,Ishak2005,KnoxSongTyson2006}) studied inconsistency between constraints on parameters from the growth rate and constraints from the expansion history to detect deviations from GR using observations. These were followed by another approach (see for example \cite{Linder2005, ZhangEtal2007, HuandSawicki2007, CaldwellCoorayandMelchiorri2007, AcquavivaEtAl2008,WangEtal2007, KunzandSapone2008, Gong2008, Daniel2008, GongIshakWang2009, BertschingerandZukin2008, HutererandLinder2007, KoivistoandMota2006, GannoujiMoraesandPolarski2009,Koyama, Zhang2006, LinderCahn, Song2007, Dore, Gabadadze, Polarski, Hu2008, Jain, Wei, Dent2009, Fu, Ishak2009, Linder2009, Serra, Song2009b, Song2009c, Thomas, Tsujikawa, Wu, GongBo2009a, GongBo2009b, Acquaviva2010, Dossett, Ferreira, Jing, Pogosian2010, Song2010}) that based their analysis on various growth parameters that take distinctive values for distinct gravity theories. For example, in GR, some parameters take the unity value, and deviation from it would indicate deviations from GR. It is worth mentioning that some useful pioneering work on the growth rate parametrization in for example \cite{Peebles1980,Fry1985,LightmanSchechter1990,WangSteinhardt1998} had
been done for a purpose other than seeking to distinguish between dark energy and modified gravity (MG). Finally, it is informative to refer to all the early work about post-GR parametrizations that had been proposed to test deviations from GR at the level of solar system tests, for example, see \cite{Will,Will2} and references therein. 

Recently, some studies used current available data sets in order to constrain the MG parameters and to see if there is an indication of any deviation from GR, but this has been in vain so far \cite{Daniel2009, Bean2010, Daniel2010, DL2010, GongBo2010, Lombriser, Toreno}.   

IIn this paper, we calculate and compare constraints and figures of merit (FoM) on MG parameters based on func- tional forms or binning methods, using combined data sets of baryon acoustic oscillations (BAO) measurements from Two-Degree Field and DR7 Sloan Digital Sky (SDSS) surveys \cite{BAOReid, Percival2009}, the WMAP 7-year (WMAP7) cosmic microwave background (CMB) temperature (TT) and polarization (TE) spectra \cite{WMAP7all}, and the supernovae Union2 compilation, which includes the 557 Type Ia SNe of the Supernovae Cosmology Project (SCP) \cite{Union2}, and references of other compiled supernovae therein, the matter power spectrum (MPK) from SDSS DR7 \cite{BAOReid}, the integrated Sachs Wolfe (ISW)-galaxy cross correlations with the 2MASS and SDSS luminous red galaxy (LRG) surveys \cite{ISWHo,ISWHirata}, and the recently refined Hubble Space Telescope (HST) Cosmic Evolution Survey (COSMOS) weak- lensing tomography analysis in \cite{Schrabback2010}.
%
%
\section{Methodology}
%
\subsection{Unmodified Growth Equations}
For convenience, we will work in the conformal Newtonian gauge, where the perturbed Friedmann- Lemaitre-Robertson-Walker metric is written as:
\be
ds^2=a(\tau)^2[-(1+2\psi)d\tau^2+(1-2\phi)dx^idx_i],
\label{eq:FLRW}
\ee
where $a(\tau)$ is the scale factor normalized to one today, the $x_i$'s are the comoving coordinates, and $\tau$ is conformal time.  $\phi$ and $\psi$ are scalar potentials describing the scalar mode of the metric perturbations.  

We can get two very useful equations that describe the evolution of the scalar potentials by using the first-order perturbed Einstein equations working in Fourier k-space.  Combining the time-time and time-space equations gives the Poisson equation.  Taking the traceless, space-space component of these equations lets us relate the two potentials.  Respectively, we have:
\bea
k^2\phi  &=&-4\pi G a^2\sum_i \rho_i \Delta_i
\label{eq:P}\\
k^2(\psi-\phi) &=& -12 \pi G a^2\sum_i \rho_i(1+w_i)\frac{\pi_i}{2},
\label{eq:2E}
\eea
where $\rho_i$ and $\pi_i$ are the density and the anisotropic stress, respectively, for matter species, $i$.  
$\Delta_i$ is the gauge-invariant, rest-frame overdensity for matter species, $i$, the evolution of which describes the growth of inhomogeneities.  It is defined by:
\be
\Delta_i = \delta_i +3\mathcal{H}\frac{q_i}{k},
\label{eq:Deltadef}
\ee 
where $\mathcal{H} =\dot{a}/a$ is the Hubble factor in conformal time, and for species $i$, $\delta_i=\delta \rho_i/\bar{\rho}$ is the fractional overdensity and $q_i$ is the heat flux and is related to the divergence of the peculiar velocity, $\theta_i$, by $\theta_i=\frac{k\ q_i}{1+w_i}$.  Enforcing the conservation of energy-momentum on the perturbed matter fluid, these quantities for uncoupled fluid species or the mass-averaged quantities for all the fluids evolve as described in \cite{Ma}:
\bea
\dot{\delta} & = & -k q +3(1+w)\dot{\phi}+3\mathcal{H}(w-c_s^2)\delta
\label{eq:deltaevo}\\
\frac{\dot{q}}{k}&= &-\mathcal{H}(1-3w)\frac{q}{k}-\frac{\dot{w}}{1+w}\frac{q}{k}+c_s^2\delta-(1+w)\frac{\pi}{2}+(1+w)\psi.
\label{eq:qevo}
\eea
Above, $w=p/\rho$ is the equation of state  and $c_s^2=\delta p/\delta \rho$ is the sound speed.  Combining these two equations, we can express the evolution of $\Delta$ by:
\be
\dot{\Delta} = 3(1+w)\dot{\phi}+3\mathcal{H}w\Delta -\left[3\mathcal{H}\frac{\dot{w}}{1+w}+k^2+3\left(\mathcal{H}^2-\dot{\mathcal{H}}\right)\right]\frac{q}{k}-3\mathcal{H}(1+w)\left(\frac{\pi}{2} +\psi\right).
\label{eq:Deltaevo}
\ee

Equations (\ref{eq:P}),(\ref{eq:2E}),(\ref{eq:deltaevo}), and (\ref{eq:qevo}) are coupled to one another, combining them, along with the evolution equations for $a(\tau)$, we can describe the growth history of structures in the universe.
%
\subsection{Modifications to the growth equations for detecting deviations from general relativity.}
Recently, a lot of attention has gone into detecting deviations from general relativity by parametrizing both modifications to Poisson's equation, (\ref{eq:P}), as well as the ratio between the two metric potentials 0 and c in the perturbed Friedmann-Lemaitre-Robertson-Walker metric (called \emph{gravitational slip} by \cite{CaldwellCoorayandMelchiorri2007}), see for example \cite{CaldwellCoorayandMelchiorri2007,Daniel2009,GongBo2010,Bean2010,Daniel2010,DL2010}.  In this paper,  {we use two continuous parametrizations and a binned parametrization that allow for modifications of  Eqs. (\ref{eq:P}) and (\ref{eq:2E}) both directly and indirectly.

\subsubsection{Examples of parametrization with functional forms for time and scale dependencies}
The first parametrization with a functional form that we use was proposed by \cite{Bean2010}. It allows modifications to Eqs. (\ref{eq:P}) and (\ref{eq:2E}) that evolve monotonically in both time and scale, and makes no assumptions as to the time when a deviation from general relativity is allowed. These modifications are as follows:
\bea
k^2\phi  &=& -4\pi G a^2\sum_i \rho_i \Delta_i \,  Q
\label{eq:ModP}\\
k^2(\psi-R\,\phi) &=& -12 \pi G  a^2\sum_i \rho_i(1+w_i)\frac{\pi_i}{2} \, Q,
\label{eq:Mod2E}
\eea
where $Q$ and $R$ are the MG  parameters.  The parameter $Q$ represents a modification to the Poisson equation, while the parameter $R$ quantifies the gravitational slip (at late times, when anisotropic stress is negligible, $R=\psi/\phi$).  These parameters are parametrized to evolve in time and scale as:
\be
X(k,a) = \left[X_0 e^{-k/k_c}+X_\infty(1-e^{-k/k_c})-1\right]a^s +1,
\label{eq:BeanEvo}
\ee
where $X$ denotes either $Q$ or $R$.  So the model parameters which can be used to detect deviations from GR are now:
$Q_0$, $R_0$, $Q_\infty$, $R_\infty$, $k_c$, and $s$.  The parameters $s$ and $k_c$ parametrize time and scale dependence respectively, with GR values $s=0$ and $k_c=\infty$.  $Q_0$ and $R_0$ are the present-day superhorizon values while $Q_\infty$ and $R_\infty$ are the present-day subhorizon values of the $Q(k,a)$ and $R(k,a)$ , all taking GR values of $1$. 

The second parametrization we use comes from \cite{GongBo2010}  {and is} used in their code \texttt{MGCAMB} \cite{MGCAMB}. Here, a modification to Eq. (\ref{eq:P}) is done indirectly by defining a modified field equation containing the parameter $\mu$, as well as, a gravitational slip parameter, $\eta$.  Explicitly these modifications are:
\bea
k^2\psi &=& -4\pi G a^2\sum_i \rho_i \Delta_i \, \mu(k,a).
\label{eq:GBmu}\\
\frac{\phi}{\psi} & = & \eta(k,a)
\label{eq:GBeta}
\eea
The parameters, $\{\mu,\,\eta\}$ are allowed to have a redshift dependence where they are  {fit} to constant values below some transition redshift, $z_s$, (note that this is not to be confused with the source galaxy distribution peak redshift) and then make a transition to a GR value of $1$ as following a hyperbolic tangent function with a transition width, $\Delta z$:
\bea 
\mu(z) &=& \frac{1-\mu_{0}}{2}\Big(1 + \tanh{\frac{z-z_s}{\Delta z}}\Big) + \mu_{0},
\label{eq:GBmuEvo}\\
\eta(z) &=& \frac{1-\eta_0}{2}\Big(1 + \tanh{\frac{z-z_s}{\Delta z}}\Big) + \eta_{0}.
\label{eq:GBetaEvo}
\eea 
Also, the parametrization uses one other parameter, $\Sigma(k,\,a)$, defined by:
\be 
\Sigma(k,a)\equiv -\frac{k^2(\psi + \phi)}{8\pi G \rho a^2 \Delta} = \frac{\mu(1+\eta)}{2}
\label{eq:SigmaMG}
\ee
but with only a redshift dependence here in this  subsection.  As we discuss in Sec. II C below, this parameter is useful to break degeneracies between the other parameters. 
\subsubsection{Example of parametrization with binning in time and scale}
Third, in order to allow for simultaneous scale (wavenumber)  and redshift (time) dependences in the $(\mu(k,a),\Sigma(k,a))$ parametrization above, we use variants of a  $2 \times 2$ binned approach (called pixillation in \cite{GongBo2010}). We use two bins for scale dependence as $0.0<k\leq k_x$ and $k_x<k<\infty$, and where we use an analysis with $k_x= 0.01\, h\, Mpc^{-1}$ and the other with $k_x= 0.1\, h\, Mpc^{-1}$.  In both cases, we bin for redshift dependence as $0<z\leq 1$ and $1<z\leq 2$. We also use a third redshift binning method with $0<z\leq 1.5$ and $1.5<z\leq 3$ and the scale bin separator at $k_x= 0.01\, h\, Mpc^{-1}$ for the data sets in this paper.  We use combinations of the parameters $\mu_{i}$ and $\Sigma_{i}$ with $i=1...4$ for each $[k,\,z]$ bins  {in a $2 \times 2$ grid as shown in Table \ref{table:Grid}.  Transitions between the} redshift bins behave as a hyperbolic tangent function with a transition width of $\Delta z = 0.05$ (almost a step), while transitions between scale bins are straightforward steps.  

\begin{center}
\begin{table}[t]
\begin{tabular}{|c|c|c|}\hline 
&\multicolumn{2}{|c|}{Redshift bins}\\\hline
Scale bins & $0.0<z\leq 1,\,1.5$ & $1,\,1.5 <z \leq 2,\,3$\\\hline
$0.0 < k \leq k_x$& $\mu_{1},\,\Sigma_{1}$& $\mu_{2},\,\Sigma_{2}$\\\hline
$k_x < k< \infty$& $\mu_{3},\,\Sigma_{3}$&$\mu_{4},\,\Sigma_{4}$ \\\hline
\end{tabular}
\caption{\label{table:Grid}
The layout of the binned parametrizations.  Specifically, for the first two binned methods this involves using $\{\mu_{1},\,\Sigma_{1}\}$ for the $0<z\leq1$ and  $0.0<k\leq k_x$ bin, $\{\mu_{2},\,\Sigma_{2}\}$ for the $1<z\leq 2$ and  $0.0<k\leq k_x$ bin, $\{\mu_{3},\,\Sigma_{3}\}$ for the $0<z\leq 1$ and  $k_x<k< \infty$ bin, and $\{\mu_{4},\,\Sigma_{4}\}$ for the $1<z\leq 2$ and  $k_x<k< \infty$ bin, and the third binned method uses $\{\mu_{1},\,\Sigma_{1}\}$ for the $0<z\leq1.5$ and  $0.0<k\leq k_x$ bin, $\{\mu_{2},\,\Sigma_{2}\}$ for the $1.5<z\leq 3$ and  $0.0<k\leq k_x$ bin, $\{\mu_{3},\,\Sigma_{3}\}$ for the $0<z\leq 1.5$ and  $k_x<k< \infty$ bin, and $\{\mu_{4},\,\Sigma_{4}\}$ for the $1.5<z\leq 3$ and  $k_x<k< \infty$ bin.}
\end{table}
\end{center}
%
\subsection{Relations between parametrizations and degeneracies among parameters}
%
%
We now turn our discussion to relating the different parametrizations to one another.  We will focus on the $\{Q,\,R\}$ and $\{\mu,\,\eta,\,\Sigma\}$ parametrizations above, as well as, a third parametrization using $\{\mathcal{G},\,\mathcal{V}\}$ used in\cite{DL2010}.  When relating the $\texttt{MGCAMB}$ and $\{\mathcal{G},\,\mathcal{V}\}$ parametrizations to that of the first, we give the relations during matter domination, assuming zero anisotropic stress.
\bea
\mu =Q R = \mathcal{V}_{}\qquad &\mbox{, }& \qquad \eta = \frac{1}{R}=2\frac{\mathcal{G}}{\mathcal{V}}-1
\label{eq:comp1}\\
\Sigma = \frac{Q(1+R)}{2} =\mathcal{G}_{}\qquad &\mbox{, }&\qquad \mu \eta = Q = 2\mathcal{G}-\mathcal{V}.
\label{eq:comp2}
\eea
We use these relations later in order to infer constraints on some parameters from the  {ones} obtained from the constraints derived from the data sets.    

As previously pointed out \cite{Bean2010,Daniel2010,DL2010,GongBo2010} some of these parameters are degenerate and it is worth introducing parameters that alleviate these degeneracies. First, the parameters $Q_{0,\infty}$ and $R_{0,\infty}$ are degenerate along a direction that we call $D_{0,\infty} = Q_{0,\infty}(1+R_{0,\infty})/2$ this degeneracy is shown with a curve in Fig. \ref{fig:BFig}. We proceed by varying $Q_{0,\infty}$ and $D_{0,\infty}$ and infer values of $R_{0,\infty}$ from the two parameters. Similarly, $\mu$ and $\eta$ are degenerate along curves of $\Sigma = \mbox{constant}$ so we vary $\mu$ and $\Sigma$ to get rid of this degeneracy. Using the parameters $D$ and $\Sigma$ is also useful because observations of the weak-lensing and ISW are sensitive to the sum of the metric potentials $\phi +\psi$ and its time derivative respectively. Thus observations are able to give us direct measurements of these parameters.    

\begin{figure}
\begin{center}
\begin{tabular}{|c|c|}
\hline

{\includegraphics[width=2.8in,height=2.in,angle=0]{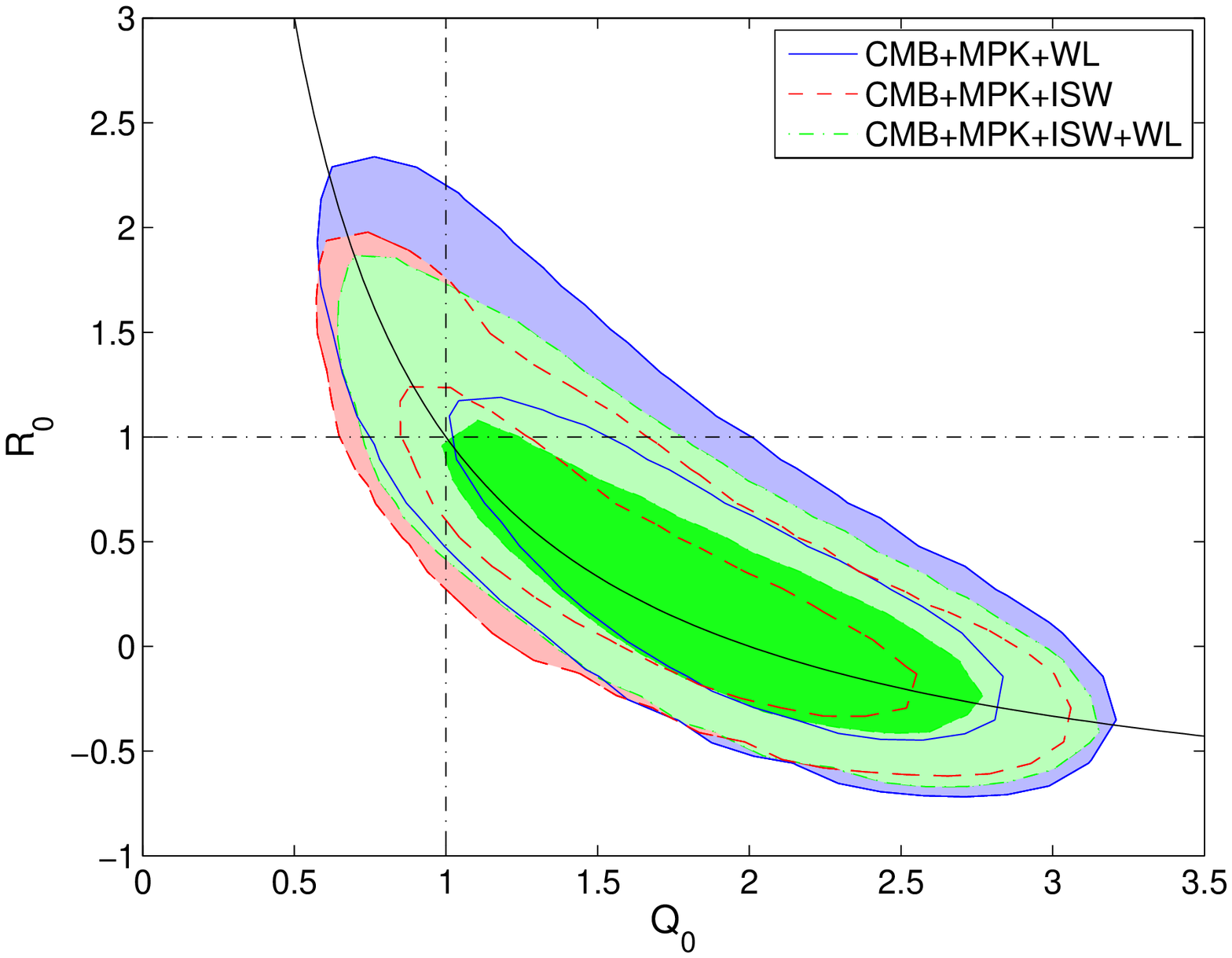}} &
{\includegraphics[width=2.8in,height=2.in,angle=0]{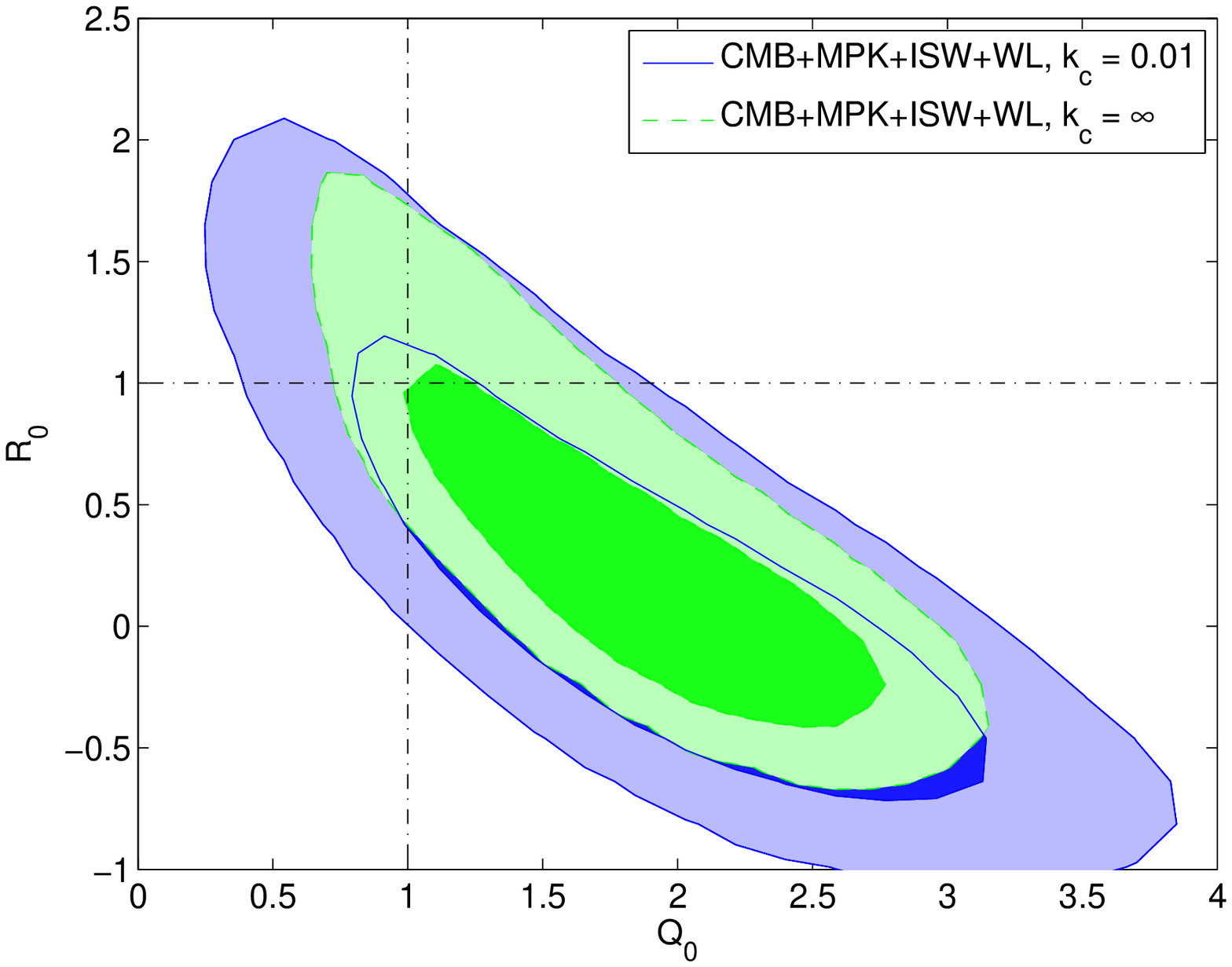}} \\
\hline  
{\includegraphics[width=2.8in,height=2.in,angle=0]{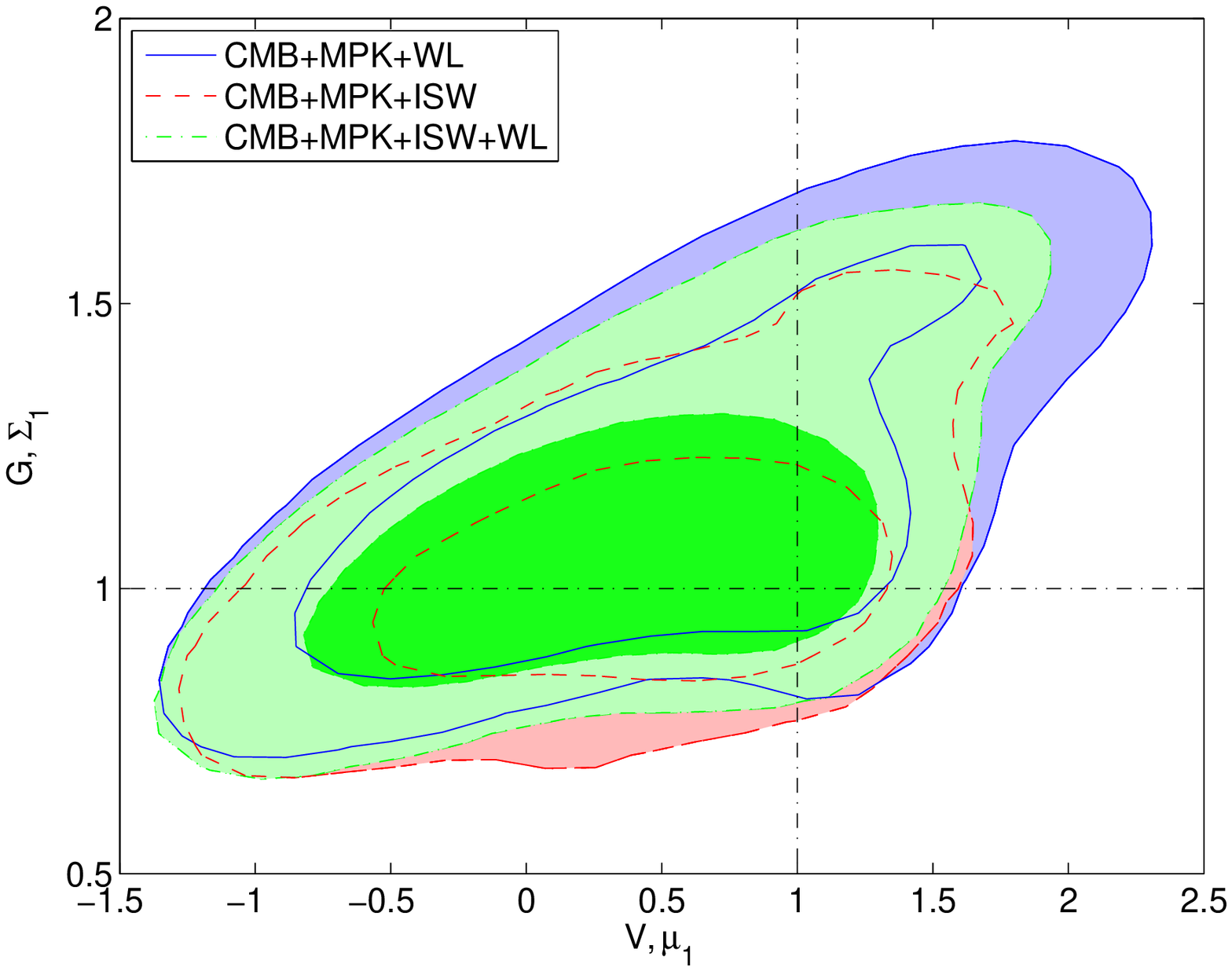}} &
{\includegraphics[width=2.8in,height=2.in,angle=0]{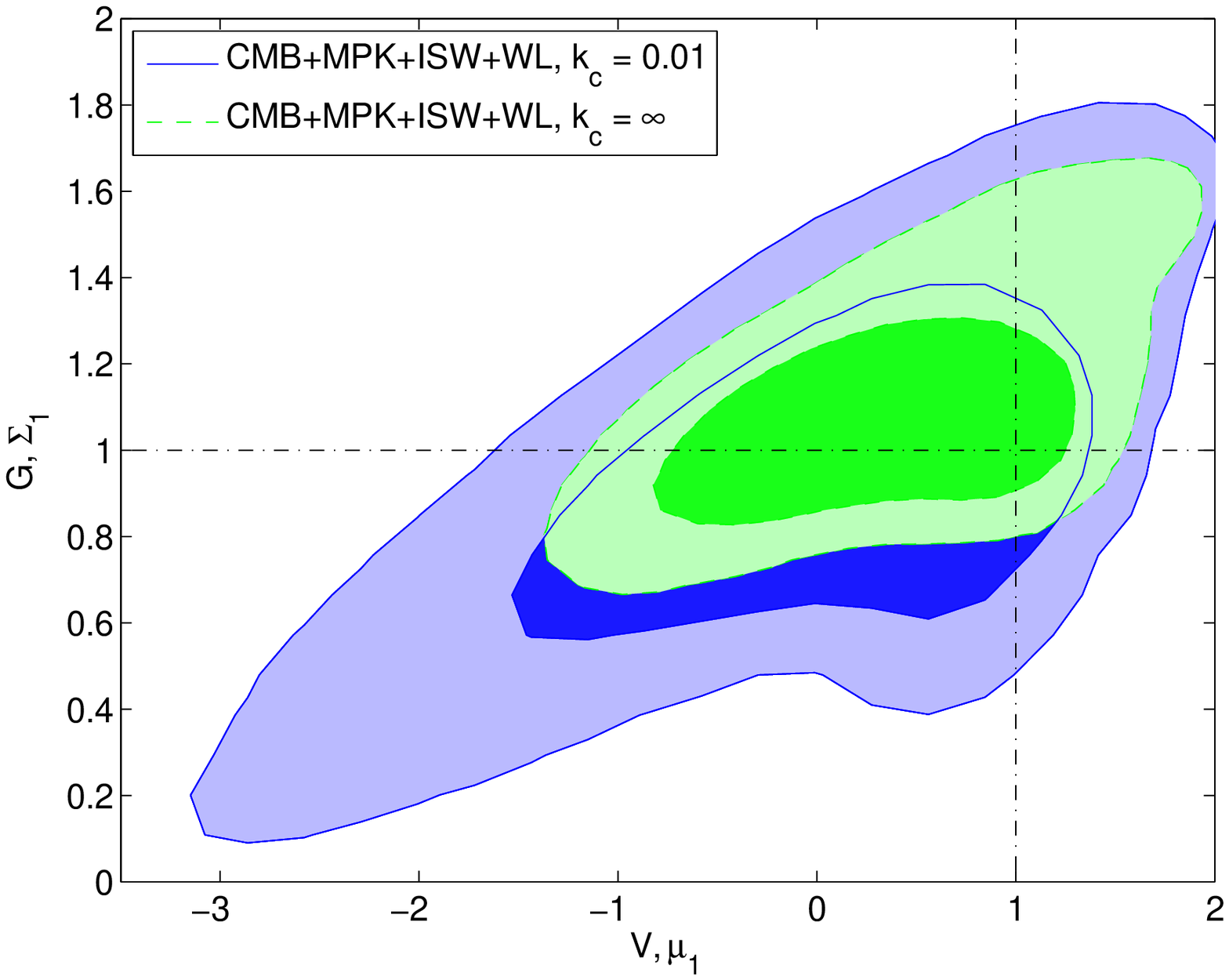}} \\
\hline  
\end{tabular}
\caption{\label{fig:BFig}
TOP: $68\%$ and $95\%$ C.L. for the constraints on $Q_0$ and $R_0$.  TOP-LEFT: Constraints with $Q$ and $R$ scale independent ($k_c = \infty$).  To show how the parameter $D_0$ actually represents the degeneracy direction of these parameters, the solid black line is the line $D_0 = 1$.   TOP-RIGHT: A comparison of constraints when $Q$ and $R$ are allowed either scale dependence ($k_c = 0.01$) or scale independence using CMB, ISW, MPK, and WL.  BOTTOM-LEFT:  Inferred $68\%$ and $95\%$ C.L. constraints for the parameters  $\mathcal{G}_{}$ or $\Sigma_1$ and $\mathcal{V}_{}$ or $\mu_1$ in the second and third parametrizations \cite{GongBo2010, DL2010}.  These constraints are inferred by using constraints on the parameters $Q_0$, $R_0$, and $D_0$ from the first parametrization \cite{Bean2010}.  For all of these constraints we set $k_c=\infty$ implying scale independence of the
parametrization (\ref{eq:BeanEvo}).  BOTTOM-RIGHT:  We compare inferred constraints on $\mathcal{G}_{}$ or $\Sigma_1$ and $\mathcal{V}_{}$ or $\mu_1$ using CMB, ISW, MPK, and WL when scale dependence is allowed in the first parametrization to constraints when the parametrization is scale independent. } 
\end{center}
\end{figure}
%
%
\subsection{Figures of merit}
%

We calculate the figures of merit for the parameter pairs used for each parametrization considered. The FoM approach has been used previously, for example, to quantify the constraining power of cosmological data on the dark-energy equation-of-state parameters, using simulated future data \cite{ReportDETF, FoMSWG, YWangFoM, MaZhangFoM, Acquaviva2010, AmaraKitching}, or current data \cite{FoM, MHH1, MHH2}. For a pair of parameters, the FoM can be defined to be proportional to the inverse of the $95\%$ confidence limit area. For more parameters, this can be generalized to be proportional to the inverse of the volumes and supervolumes of the (super)ellipsoids, again representing the confidence limits in these parameter hyperspaces. A variety of choices for the constants of proportionality has been made in previous work and we simply chose to set it to one. Assuming a Gaussian distribution, the FoM for two parameters is inversely proportional to the area of the ellipse comprising the contour of constraints for the two parameters and so we define our FoM  \cite{FoM},
\be 
\mbox{FoM}=(det\, C)^{-1/2}
\label{eq:FoM}
\ee
where $C$ is the 2D covariance matrix for the two parameters \cite{FoMWang}. This is similar to the FoM used by the Dark Energy Task Force for the equation of state parameters. 
Obviously, tighter constraints lead to higher FoMs. 

For two uncorrelated parameters with Gaussian uncertainties, the FoM does not provide any more information than the individual uncertainties in describing the constraining power of the data sets \cite{ReportDETF}, but our results are not purely Gaussian, and our parameters for which we show FoM's are not always uncorrelated.  The  {FoM's} are very informative to tell us which data set is most significant in constraining the parameters and in our study here, were useful in indicating some tension between the data sets.  It is of course understood that the FoM as we have defined it here, is only an approximation with non-Gaussian uncorrelated parameters, but accurate within $\sim 10\%$ in the worst cases \cite{FoM}.  The FoM is less accurate for two degenerate parameters, but still provides a good approximation to compare the constraining power of data sets.  In fact, at first glance, when looking at the constraints in the tables and figures below, without the FoM, it is more difficult to determine which combination of data sets is more sensitive because at times there is a shift in the allowed parameter space for the different combinations of data sets.

This metric is useful in comparing the constraining power of various combinations of data sets that we used, namely WMAP7 temperature and polarization spectrum (CMB), the matter power spectrum (MPK), ISW-galaxy cross correlations (ISW) and HST COSMOS weak-lensing tomography (WL).  In each case, we use Union2 supernovae (SN), BAO, BBN, AGE and $H_0$, as well.  

\section{data sets for observational constraints}
%
We use Type Ia SNe from the Union2 compilation\cite{Union2}, BAO from Two-Degree Field and SDSS DR7 \cite{BAOReid, Percival2009}, and WMAP7 spectrum.  The WMAP7 data, MPK from SDSS DR7 \cite{BAOReid}, the ISW-galaxy cross correlations \cite{ISWHo,ISWHirata} and COSMOS weak-lensing tomography \cite{Schrabback2010} will constrain the growth as described below.
We compare to the full temperature (TT) and polarization (TE) spectra using the WMAP 7-year data release using the likelihood  routine provided by the WMAP team \cite{WMAP7all}.
Next, for the BAO, we follow the most recent comparison \cite{BAOReid, Percival2009} and define the ratio of the sound horizon, $r_s(z_d)$ to the effective distance, $D_V$ as a fit for SDSS.
We also add the prior $H_0=74.2\pm 3.6$ km/s/Mpc given by \cite{riess}, and the priors on the age of the Universe (AGE) $10\,$Gyrs$<$AGE$<20\,$Gyrs, and $\Omega_b h^2 = 0.022 \pm 0.002$ from big bang nucleosynthesis (BBN) . 
\begin{figure}
\begin{center}
\begin{tabular}{|c|c|}
\hline
{\includegraphics[width=2.8in,height=2.in,angle=0]{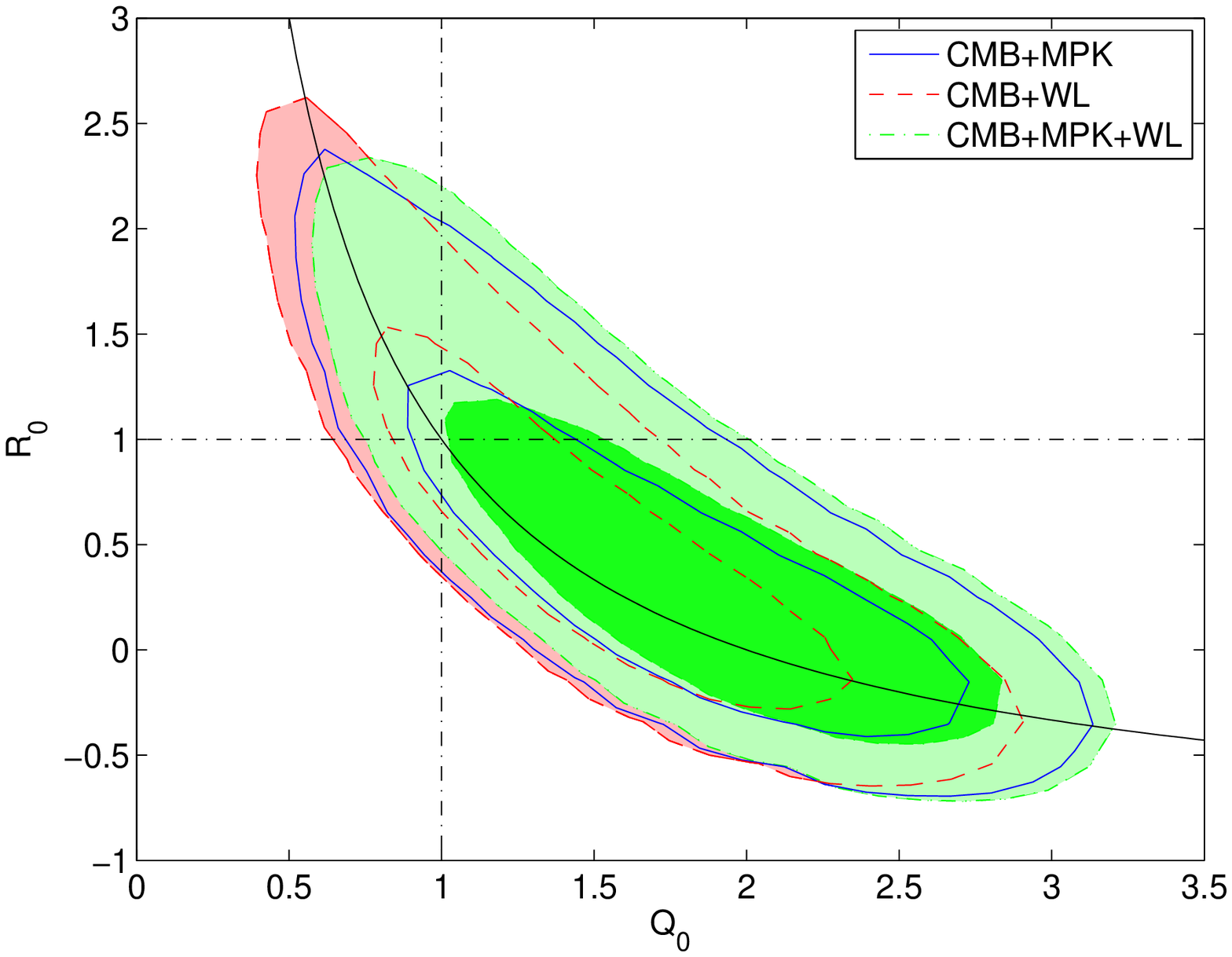}}&
{\includegraphics[width=2.8in,height=2.in,angle=0]{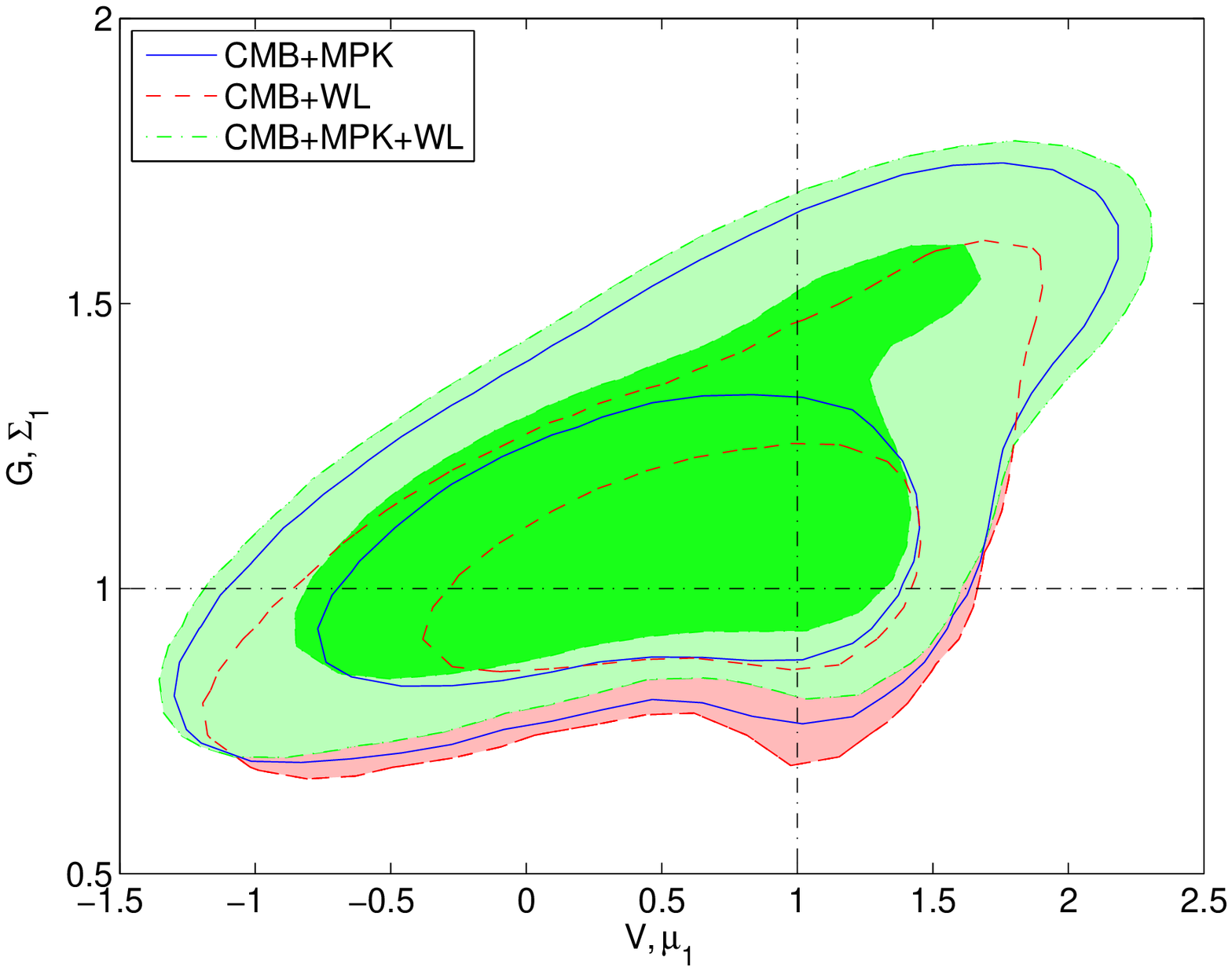}} \\
\hline  
\end{tabular}
\caption{\label{fig:MPKWLTENS}
We compare the $68\%$ and $95\%$ C.L. constraints on the $Q_0$ and $R_0$ (LEFT) as well as the inferred parameters  $\mathcal{G}_{}$ or $\Sigma_1$ and
$\mathcal{V}_{}$ or $\mu_1$ (RIGHT) for CMB+MPK (solid line) and CMB+WL (dashed line) and CMB+MPK+WL (dash-dotted) to show the tension between these data sets.   Notice the combined constraints are outside those of the individual data sets.
} 
\end{center}
\end{figure}
A possible deviation from GR is allowed by the modified Poisson equation and inequality between $\phi$ and $\psi$.  The growth of structure is important in constraining the model parameters for a deviation from GR in this modified description.  
For the total matter power spectrum today, for example, see \cite{Bean2010}, we use
\be 
P(k) = 2\pi^2\Big(\frac{k}{h}\Big)^3 \Delta_{tot}^2,
\label{eq:MPKtoday}
\ee
where $\Delta_{tot} = (\rho_c \Delta_c + \rho_b \Delta_b)/(\rho_c+\rho_b)$ is the gauge-invariant, rest-frame, overdensity for cold dark matter  and baryons combined.  Recall, that the matter power spectrum is sensitive to the growth of structure by constraining $\mu$ through $\Delta$ \cite{DL2010, GongBo2010,Bean2010}.  WMAP7 data also provides large scale ISW measurements.  The ISW effect describes the evolution in time of the gravitational potentials.  The photons from the CMB gain energy falling into varying potential wells, and may not return it all on exit.  This causes a secondary anisotropy of the CMB and can be analyzed directly with the derivatives of $\phi$ and $\psi$ to give an indication of deviation from GR, as described in  \cite{DL2010, GongBo2010,Bean2010}.  It is difficult to distinguish these CMB temperature anisotropies signatures from rest of CMB anisotropies because it has $1/10$ the amplitude, so we need to correlate this effect to the large scale structure of the universe.  For this we use the ISW-galaxy cross correlations which include the use of 2MASS and SDSS LRG galaxy autocorrelations explored in \cite{ISWHo, ISWHirata} and recently used by \cite{Bean2010,DL2010}.  

We further constrain the parameters with expansion history and the effect of the growth of structure from the weak-lensing tomography shear-shear cross correlations of the HST COSMOS survey recently compiled by Schrabback et. al. \cite{Schrabback2010}.  The weak-lensing constrains the growth of structure through the combination of $\phi$ and $\psi$ as described in  \cite{DL2010, GongBo2010,Bean2010}.  The authors of \cite{Schrabback2010} perform a refined analysis of the HST COSMOS survey of \cite{Scoville2007}, in combination with the COSMOS-30 photometric redshift catalogue provided by \cite{Ilbert2009}.  The shear-shear cross correlations were calculated between 6 redshift bins, $0.0<z<0.6$, $0.6<z<1.0$, $1.0<z<1.3$, $1.3<z<2.0$, $2.0<z<4.0$ and a sixth bin that contains all faint galaxies with a numerically estimated redshift distribution from $0.0<z<5.0$, see Fig. 6 of \cite{Schrabback2010}.  Certain exclusions were made as described in \cite{Schrabback2010}, such as the LRGs to avoid G-I intrinsic alignment bias and the lowest angular theta bin because of model uncertainties.  Only bright galaxies ($i<24$) were used in the first bin $(z<0.6)$. Also, autocorrelations were not used in bins $1-5$ to reduce the effect of I-I intrinsic alignments. For comparing the  shear-shear cross correlations, we modify the publicly available code for the COSMOS 3D weak-lensing built by Lesgourgues et. al. \cite{Lesgourgues} to incorporate the shear cross correlations as seen in, for example, \cite{Schrabback2010} (see endnote \cite{jnd}).

For the ISW-galaxy cross correlations we modified our code as described in previous works \cite{GongBo2010,Bean2010,DL2010} and refer the readers to those works for an extended discussion of the implications of these modifications.  Here, however, we use the newly refined HST-COSMOS 3D weak-lensing tomography by \cite{Schrabback2010} and describe the modeling of this data below (We note that during the preparation of this manuscript, \cite{Toreno} released a paper fitting MG  parameters using this data set as well).  

In order to model the weak-lensing, the shear cross correlation functions $ \xi^{kl}_{+,-}(\theta) $ between bins $k,l$ are given by:
\be 
 \xi^{kl}_{+,-}(\theta) = \frac{1}{2\pi} \int^{\infty}_{0} d\ell\ \ell\  J_{0,4}(\ell\theta)P^{kl}_{\kappa} (\ell),
 \label{eq:ShearCrossCorrelations}
 \ee
 where $J_n$ is the $n^{th}$-order Bessel function of the first kind, $\ell$ is the modulus of the two-dimensional wave vector, and $P^{kl}_{\kappa}$ is the convergence cross-power spectra between bins $k,l$.  These cross-power spectra are computed from integrating the three-dimensional (nonlinear) power spectrum $P_{\delta}$,
\be 
P^{kl}_{\kappa}(\ell) = \frac{9H^4_0 \Omega_m^2}{4c^4} \int^{\chi_h}_{0}  d\chi \frac{g_k(\chi)g_l(\chi)}{a^2(\chi)}P_{\delta} \Big(\frac{\ell}{f_K(\chi)},\chi \Big),
\label{eq:3DPowerSpectrum}
\ee
with comoving radial distance $\chi$, comoving distance to the horizon $\chi_h$, and comoving angular diameter distance $f_K(\chi)$.  We note here that the effect of the MG  parameters on the input power spectrum from the transfer function has been implemented in our weak-lensing code, while the effect of the growth has been accounted  {for} by our modifications to \texttt{CAMB} \cite{LewisCAMB}, and both are included in $P_\delta$ of Eq. (\ref{eq:3DPowerSpectrum}).
Also, following \cite{Schrabback2010}, we weight the geometric lens-efficiency factors
\be 
g_k(\chi)  \equiv \int^{\chi_h}_{\chi} d\chi' p_k(\chi') \frac{f_K(\chi'-\chi)}{f_K(\chi')},
\label{eq:LensEfficiency}
\ee
corresponding to the redshift distributions $p_k$ for the two compared redshift bins.
We use the $160 \times 160$ dimensional covariance matrix provided by Schrabback et. al. \cite{Schrabback2010} and for the corresponding weak shear correlations. We also apply the correction to the inverse covariance, $Cov^{-1}$, seen in \cite{Schrabback2010, Hartlap2007}, as
\be 
\stackrel{\ast}{Cov}{}^{-1} = 0.4390 \,Cov^{-1}.
\label{eq:InMatrix}
\ee
 for $288$ independent realizations and a $160$ dimensional data vector.  Following instruction from \cite{Schrabback2010}, we weight the predictions for the $k^{th}$ $\theta$ bin, $\theta_{k}$, with logarithmic spaced upper and lower limits $\theta_{k,max}$ and $\theta_{k,min}$, respectively,  based on the number of galaxy pairs,$N$ for a given $\theta$ as
 \be 
 \xi(\theta_{k})= \frac{\int^{\theta_{k,max}}_{\theta_{k,min}} N(\theta') \xi_{+,-}(\theta') d\theta'}{\int^{\theta_{k,max}}_{\theta_{k,min}} N(\theta') d\theta'},
 \label{eq:WeightTheta}
 \ee
 with $N(\theta) \propto \theta(0.0004664+\theta(0.0044118-8.90878\times 10^{-5}\theta))$ where the units of $\theta$ is arcminutes.  \cite{Schrabback2010,Schrabback2010private_comm}.  Finally, the $10\%$ uncertainty in the numerical estimate of the galaxy redshift distribution of bin 6 is applied, following \cite{Schrabback2010}, as $p_6(z,f_z) \equiv p_6(f_z z)$, where $f_z$ is a nuisance parameter marginalized in the results below.

\begin{center}
\begin{table}[t]
\begin{tabular}{|c|c|c|c|c|c|c|c|c|}\hline
\multicolumn{9}{|c|}{\bfseries Constraints for the parameters $\{Q_{0,\infty},\,R_{0,\infty},\,D_{0,\infty}$\}}\\ \hline
\multicolumn{9}{|c|}{$k_c = \infty$}\\ \hline
data set&$\mbox{FOM}_{Q_0;R_0}$&$\mbox{FOM}_{Q_0;D_0}$ & $Q_0$&$R_0$&$D_0$&$Q_\infty$&$R_\infty$&$D_\infty$\\
&&&&(derived)&&&(derived)&\\ \hline
CMB, MPK&5.53 &8.81 &$[ 0.69,\,2.79]$ &$[ -0.34,\,2.04]$ &$[ 0.79,\,1.62]$ & -- &-- &--\\ \hline
CMB, ISW &6.15&9.80&$[ 0.70,\,2.76]$ &$[ -0.33,\,1.85]$ &$[ 0.79,\,1.55]$ & -- &-- &-- \\ \hline
CMB, WL &6.44&12.15&$[ 0.57,\,2.54]$ &$[ -0.30,\,2.24]$ &$[ 0.76,\,1.47]$ & -- &-- &-- \\ \hline
CMB, ISW, WL &8.46  &14.96&$[ 0.67,\,2.53]$ &$[ -0.33,\,1.84]$ &$[ 0.75,\,1.34]$ & -- &-- &--\\ \hline 
CMB, MPK, ISW &7.53 &11.97&$[ 0.80,\,2.73]$ &$[ -0.35,\,1.62]$ &$[ 0.76,\,1.46]$ & -- &-- &-- \\ \hline 
CMB, MPK, WL &5.08 &7.96&$[ 0.73,\,2.80]$ &$[ -0.34,\,2.13]$ &$[ 0.82,\,1.65]$ & -- &-- &--\\ \hline
CMB, MPK, ISW, WL &7.09 &10.06&$[ 0.83,\,2.82]$ &$[ -0.37,\,1.60]$ &$[ 0.77,\,1.56]$ & -- &-- &-- \\ \hline \hline
\multicolumn{9}{|c|}{$k_c = 0.01$}\\ \hline
data set &$\mbox{FOM}_{Q_0;R_0}$&$\mbox{FOM}_{Q_0;D_0}$& $Q_0$&$R_0$&$D_0$&$Q_\infty$&$R_\infty$&$D_\infty$\\
&&&&(derived)&&&(derived)&\\ \hline
CMB, MPK, ISW, WL&4.07  &4.53 &$[ 0.57,\,3.35]$ &$[ -0.79,\,1.67]$ &$[ 0.29,\,1.60]$&$[ 0.46,\,4.01]$ &$[-0.82,\,2.28]$ &$[ 0.26,\,2.00]$\\
\hline 
\end{tabular}
\begin{tabular}{|c|c|c|c|}\hline
\multicolumn{4}{|c|}{}\\ \hline
\multicolumn{4}{|c|}{\bfseries Inferred parameter constraints on the parameters  $\mathcal{G}_{},\,\Sigma_1$ and $\mathcal{V}_{},\,\mu_1$ from parameters $Q_0$, $R_0$, and $D_0$}\\ \hline 
\multicolumn{4}{|c|}{$k_c = \infty$}\\ \hline
data set &$\mbox{FOM}_{\mathcal{G};\mathcal{V}}$ & $ \qquad \qquad \mathcal{G}_{}$,$\Sigma_1\qquad \qquad $&$\qquad \qquad\mathcal{V}_{}$,$\mu_1\qquad \qquad$\\ \hline
CMB, MPK&8.81 &$[ 0.79,\,1.62]$ &$[ -0.90,\,1.79]$  \\ \hline
CMB, ISW &9.80&$[ 0.79,\,1.55]$ &$[ -0.86,\,1.57]$  \\ \hline
CMB, WL &12.15&$[ 0.76,\,1.47]$ &$[ -0.72,\,1.60]$  \\ \hline 
CMB, ISW, WL &14.95 &$[ 0.75,\,1.34]$ &$[ -0.78,\,1.38]$ \\ \hline
CMB, MPK, ISW &11.97&$[ 0.76,\,1.46]$ &$[ -0.89,\,1.42]$   \\ \hline 
CMB, MPK, WL &7.96 &$[ 0.82,\,1.65]$ &$[ -0.90,\,1.91]$ \\ \hline
$\qquad$CMB, MPK, ISW, WL$\qquad$ &10.06&$[ 0.77,\,1.56]$ &$[ -0.98,\,1.57]$  \\ \hline \hline
\multicolumn{4}{|c|}{$k_c = 0.01$}\\ \hline
data set &$\mbox{FOM}_{\mathcal{G};\mathcal{V}}$ & $ \qquad \qquad \mathcal{G}_{}$,$\Sigma_1\qquad \qquad $&$\qquad \qquad\mathcal{V}_{}$,$\mu_1\qquad \qquad$\\ \hline
CMB, MPK, ISW, WL&4.53 &$[ 0.29,\,1.60]$&$[ -2.46,\,1.48]$ \\ \hline
\end{tabular}
\caption{\label{table:B1}
TOP: $95\%$ C.L. for the parameters $Q_0$, $R_0$,$D_0$, $Q_\infty$, $R_\infty$, and $D_\infty$ in the first parametrization \cite{Bean2010}.  For these constraints, we marginalize over the parameter $s$ which characterizes the redshift (time) dependence of the parametrization.\\
BOTTOM: Inferred $95\%$ C.L. constraints for the parameters  $\mathcal{G}$ or $\Sigma_1$ and $\mathcal{V}$ or $\mu_1$ in the second and third parametrizations \cite{GongBo2010, DL2010}.  These constraints are inferred by using constraints on the parameters $Q_0$, $R_0$, and $D_0$ from the first parametrization \cite{Bean2010}.     
}
\end{table}
\end{center}
%
\section{Results and discussion}
%
%
%
\begin{center}
\begin{table}[t]
\resizebox{7.0in}{!} {
\begin{tabular}{|c|c|c|c|c|c|c|c|c|}\hline
\multicolumn{9}{|c|}{\bfseries Constraints on   $\mu_0$, $\eta_0$, and $\Sigma_0$ for redshift dependence}\\ \hline
&\multicolumn{4}{|c|}{$z_s=1$}&\multicolumn{4}{|c|}{$z_s=2$}\\ \hline
\multicolumn{9}{|c|}{$\Delta z = 0.05$ }\\ \hline
data set&FoM&$\mu_0$&$\Sigma_0$&$\eta_0$&FoM&$\mu_0$&$\Sigma_0$&$\eta_0$\\
&&&&(derived)&&&&(derived)\\ \hline 
CMB, WL &94.61& $[ 0.621,\,1.622]$ &$[ 0.948,\,1.100]$& $[ 0.233,\,2.328]$&162.5&$[ 0.645,\,1.351]$ & $[ 0.946,\,1.080]$&$[ 0.467,\,2.206]$\\ \hline 
CMB, MPK &100.9&$[ 0.618,\,1.557]$ &$[ 0.955,\,1.103]$&$[ 0.297,\,2.350]$&177.0 &$[ 0.618,\,1.298]$ & $[ 0.955,\,1.079]$&$[ 0.542,\,2.327]$ \\ \hline 
CMB, ISW &114.2&$[ 0.590,\,1.427]$ &$[ 0.954,\,1.105]$ &$[ 0.415,\,2.505]$&196.0&$[ 0.635,\,1.228]$ & $[ 0.952,\,1.082]$ &$[ 0.637,\,2.251]$\\ \hline 
CMB, MPK, WL &104.5&  $[ 0.624,\,1.549]$ &$[ 0.960,\,1.108]$ &$[ 0.317,\,2.334]$&179.4&$[ 0.643,\,1.331]$ & $[ 0.959,\,1.082]$ &$[ 0.514,\,2.221]$\\ \hline 
CMB, ISW, WL &123.9& $[ 0.589,\,1.399]$ &$[ 0.954,\,1.098]$ &$[ 0.438,\,2.499]$&209.2&$[ 0.627,\,1.198]$ & $[ 0.954,\,1.081]$&$[ 0.658,\,2.297]$ \\ \hline 
CMB, MPK, ISW &132.9& $[ 0.594,\,1.355]$ &$[ 0.956,\,1.098]$ &$[ 0.487,\,2.472]$&230.2&$[ 0.608,\,1.157]$ & $[ 0.957,\,1.076]$&$[ 0.723,\,2.372]$ \\ \hline
CMB, MPK, ISW, WL &132.5& $[ 0.592,\,1.345]$ &$[ 0.961,\,1.106]$ &$[ 0.520,\,2.483]$&229.1&$[ 0.626,\,1.182]$ & $[ 0.962,\,1.081]$  &$[ 0.706,\,2.307]$\\ \hline \hline
\multicolumn{9}{|c|}{ Marginalizing over $\Delta z$ }\\ \hline
data set&FoM&$\mu_0$&$\Sigma_0$&$\eta_0$&FoM&$\mu_0$&$\Sigma_0$&$\eta_0$\\
&&&&(derived)&&&&(derived)\\ \hline 
CMB, MPK, ISW, WL &74.29& $[ 0.618,\,1.371]$ &$[ 0.963,\,1.226]$&$[ 0.527,\,2.471]$ &194.0&$[ 0.674,\,1.176]$ & $[ 0.957,\,1.113]$&$[ 0.706,\,2.089]$ \\ \hline
\hline 
&\multicolumn{4}{|c|}{$z_s=3$}&\multicolumn{4}{|c|}{$z_s=0.76$}\\ \hline
\multicolumn{9}{|c|}{$\Delta z = 0.05$ }\\ \hline
data set&FoM&$\mu_0$&$\Sigma_0$&$\eta_0$&FoM&$\mu_0$&$\Sigma_0$&$\eta_0$\\ \hline 
CMB, WL &223.9&$[ 0.709,\,1.249]$ & $[0.944,\,1.075]$&$[0.578,\,1.906]$&75.74&$[ 0.631,\,1.676]$ & $[0.944,\,1.119]$&$[0.209,\,2.276]$\\ \hline 
CMB, MPK &231.4&$[0.678,\,1.214]$ & $[0.955,\,1.076]$&$[0.637,\,2.063]$&83.27&$[ 0.629,\,1.622]$ & $[0.952,\,1.122]$&$[0.254,\,2.281]$\\ \hline 
CMB, ISW &250.9&$[0.724,\,1.179]$ & $[0.949,\,1.082]$&$[0.684,\,1.861]$&89.01&$[ 0.606,\,1.531]$ & $[0.949,\,1.121]$&$[0.331,\,2.410]$\\ \hline 
CMB, MPK, WL&225.8&$[ 0.709,\,1.246]$ & $[ 0.952,\,1.077]$&$[ 0.602,\,1.911]$&83.35&$[ 0.629,\,1.625]$ & $[0.955,\,1.124]$&$[0.258,\,2.302]$\\ \hline 
CMB, ISW, WL&288.1&$[0.717,\,1.146]$ & $[0.951,\,1.076]$&$[0.721,\,1.881]$&87.78&$[ 0.599,\,1.539]$ & $[0.944,\,1.115]$&$[0.313,\,2.424]$\\ \hline 
CMB, MPK, ISW &296.6&$[ 0.681,\,1.109]$ & $[ 0.954,\,1.077]$&$[ 0.797,\,2.045]$&96.48&$[ 0.600,\,1.488]$ & $[0.948,\,1.115]$&$[0.358,\,2.428]$\\ \hline
CMB, MPK, ISW, WL&316.9&$[ 0.703,\,1.127]$ & $[ 0.961,\,1.076]$&$[ 0.769,\,1.961]$&95.59&$[ 0.603,\,1.492]$ & $[0.955,\,1.121]$&$[0.368,\,2.423]$\\ \hline \hline
\multicolumn{9}{|c|}{ Marginalizing over $\Delta z$ }\\ \hline
data set&FoM&$\mu_0$&$\Sigma_0$&$\eta_0$&FoM&$\mu_0$&$\Sigma_0$&$\eta_0$\\ \hline
CMB, MPK, ISW, WL&272.8 &$[ 0.729,\,1.139]$ & $[ 0.956,\,1.095]$&$[ 0.758,\,1.859]$&43.15&$[ 0.589,\,1.559]$ & $[0.948,\,1.298]$&$[0.361,\,2.689]$\\ \hline
\end{tabular}
}
\caption{\label{table:obs}
 The FoM for each data set and $\{\mu_0,\,\Sigma_0\}$ is given.$95\%$ C.L. for the parameters $\mu_0$, $\eta_0$, and $\Sigma_0$ from  { the second parametrization \cite{GongBo2010,MGCAMB}} for redshift dependence $z_s=1,\,2,\,3$ for different combinations of data sets. We also add the physically motivated transition redshift (from deceleration to acceleration in an Lambda cold dark matter model) given by $1+z_{trans}=({2\Omega_{\Lambda}}/{\Omega_m})^{1/3}\approx 1.76$ (for example see Eq. (14) in \cite{IshakReport}) for comparison. The data used is described in the text above and indicated as: WMAP7 temperature and polarization spectrum (CMB), the matter power spectrum (MPK), ISW-galaxy cross correlations (ISW), weak-lensing tomography (WL). In all cases, the data used is combined with SN, BAO, BBN, AGE and $H_0$. 
}
\end{table}
\end{center}

Monte Carlo Markov chains (MCMC's) are used to compute the likelihoods for 
the parameters in the model.  This method randomly chooses values for the above parameters and, based on the $\chi^2$ obtained, either accepts or rejects the 
set of parameters via the Metropolis-Hastings algorithm.  When a set of 
parameters is accepted it is added to the chain and forms a new starting point 
for the next step.  The process is repeated until the specified convergence 
is reached (via the Raftery-Lewis convergence criterion). We perform the best-fit of the parameters in the various models by $\chi^2$ minimization, i.e. minimizing for example $\chi^2=\chi^2_{SN}+\chi^2_{BAO}+\chi^2_{CMB}+\chi^2_{ISW}+\chi^2_{WL}+\chi^2_{MPK}$ via a maximum likelihood analysis. 
To get best-fits for the MG  parameters of a given parametrization we use, as discussed above modified versions of the publicly available ISW and Weak-lensing Likelihood code \cite{ISWHo}, HST-COSMOS weak-lensing code \cite{Lesgourgues}, \texttt{CAMB} \cite{LewisCAMB}, and \texttt{CosmoMC} \cite{cosmomc}.  For these modified codes, see \cite{jnd}. We fix the dark energy equation of state $w = -1$ assuming a $\Lambda$CDM expansion. Then in addition to varying the MG  parameters for a given parametrization, we vary the six core cosmological parameters:  $\Omega_bh^2$ and the $\Omega_c h^2$, the baryon and cold-dark matter physical density parameters, respectively; $\theta$, the ratio of the sound horizon to the angular diameter distance of the surface of last scattering; $\tau_{rei}$, the reionization optical depth; $n_s$, the spectral index; and $\ln10^{10} A_s$, the amplitude of the primordial power spectrum.  These parameters are allowed to vary within the wide range of values that are defaulted in the \texttt{params.ini} file of \texttt{CosmoMC}.  Additionally, as discussed in the previous section, when using the WL data set we marginalize over the nuisance parameter $f_z$. We provide various $95\%$ confidence limits on the MG  parameters along with the corresponding contour plots. We also calculate the figure of merit (FoM) for the parameter combinations for the parametrization considered. 

As described above,for the first parametrization, i.e. Eqs. (\ref{eq:ModP}) and (\ref{eq:Mod2E}), we follow the technique of varying $Q_{0,\infty}$ and $D_{0,\infty} = Q_{0,\infty}\left(1+R_{0,\infty}\right)/2$,  while inferring values of $R_{0,\infty}$ from the other two parameters.   We allow these parameters to vary $0<Q_{0,\infty}<10$, $0<D_{0,\infty}<10$, additionally ruling out parameter combinations where $R_{0,\infty}<-1$.  We fit both scale-independent and scale-dependent forms of this parametrization, in both cases we allow for time dependence by marginalizing over the parameter $s$ as described earlier. For scale-independent modifications, we fix $k_c = \infty$ and of course vary only the parameters $Q_0$ and $D_0$, while for scale-dependent modifications we fix $k_c = 0.01$, allowing all four parameters $Q_{0,\infty}$ and $D_{0,\infty}$ to vary.  The $95\%$ C.L.'s on these parameters and corresponding FoM's for various combinations of cosmological probes are given in Table \ref{table:B1}.  We also plot the $68\%$ and $95\%$ 2D contours for $Q_0$ and $R_0$ in Fig. \ref{fig:BFig}.  We find in all cases consistency with general relativity at the $95\%$ confidence level.

We note here that we do not find the slight $2\sigma$ deviation from GR on the parameter $Q_0$ seen in \cite{Bean2010} when combining all the data sets. Rather in all cases for this parametrization we find parameter values consistent with general relativity at the $95\%$ confidence level and a little looser than the results given in \cite{Bean2010}. We were however able to reproduce almost exactly the same results of \cite{Bean2010} but only when fixing the core cosmological parameters to their WMAP7 \cite{WMAP7all} best-fit values and allowing only the MG  parameters to vary.  We believe that the method of varying all the cosmological parameters simultaneously with the MG  parameters gives more realistic and conservative results and those are the results that we reported in our Table \ref{table:B1} and Figs. \ref{fig:BFig} and \ref{fig:MPKWLTENS}

Next, using the constraints we attained on the parameters of the first parametrization \cite{Bean2010} we can, with the help of Eqs. (\ref{eq:comp1}) and (\ref{eq:comp2}), infer constraints on the low redshift large scale parameters ($\mu_1$ and $\Sigma_1$ below) and ($\mathcal{G}_{}$ and $\mathcal{V}_{}$ in the low redshift and small $k-$ bin therein).  We list these inferred $95\%$ CL's and their FoM's in the second part of Table \ref{table:B1} and plot the $68\%$ and $95\%$ 2D contours in Fig. \ref{fig:BFig}.  We verified that the shapes of the contours plotted for the inferred parameters reflect the physical priors placed on the noninferred parameters as discussed above. Again, these results show consistency with general relativity at the $95\%$ confidence level.

\begin{figure}
\begin{center}
\begin{tabular}{|c|c|c}
\hline 
{\includegraphics[width=2.in,height=2.in,angle=0]{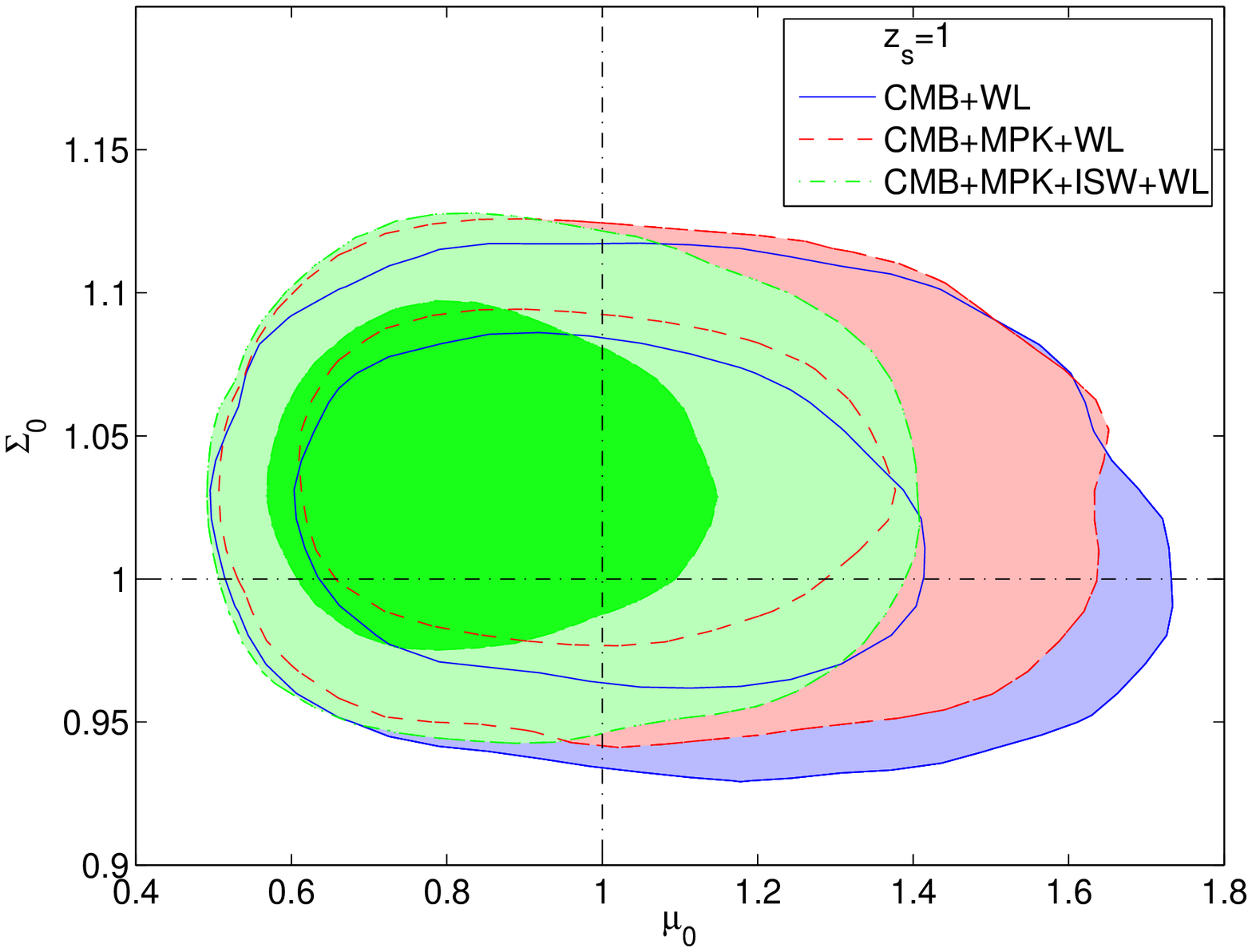}} &
{\includegraphics[width=2.in,height=2.in,angle=0]{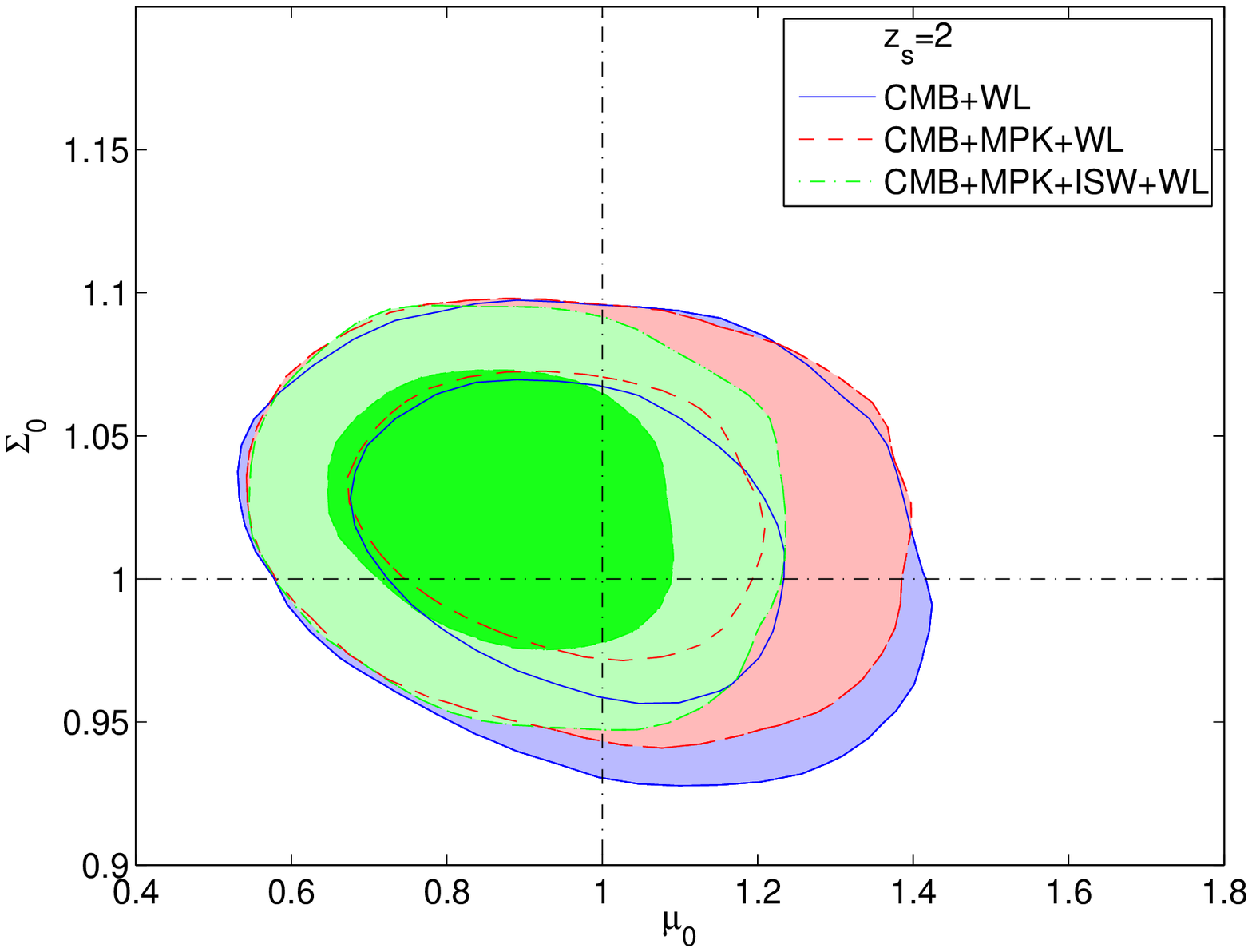}} &
{\includegraphics[width=2.in,height=2.in,angle=0]{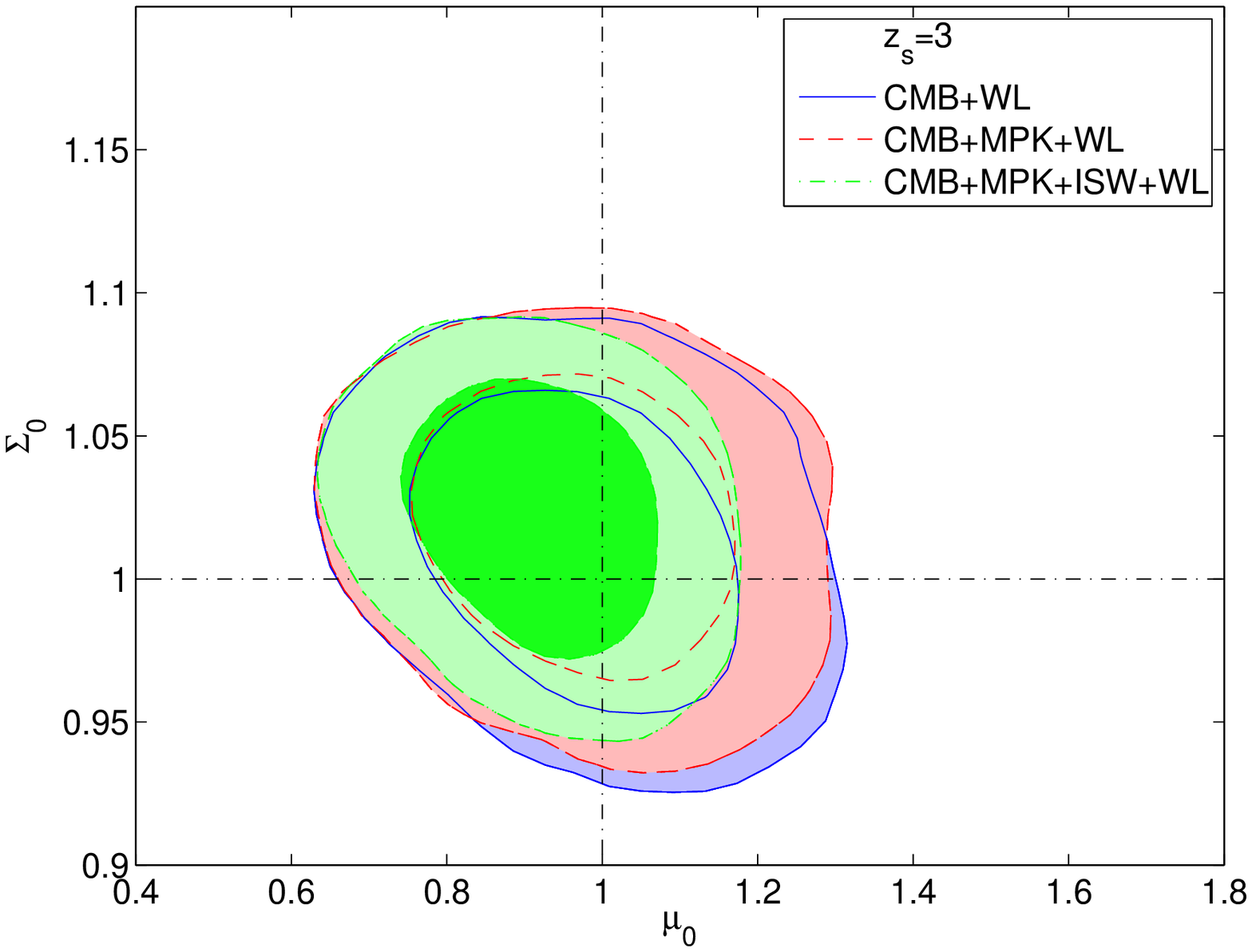}} \\
\hline  
{\includegraphics[width=2.in,height=2.in,angle=0]{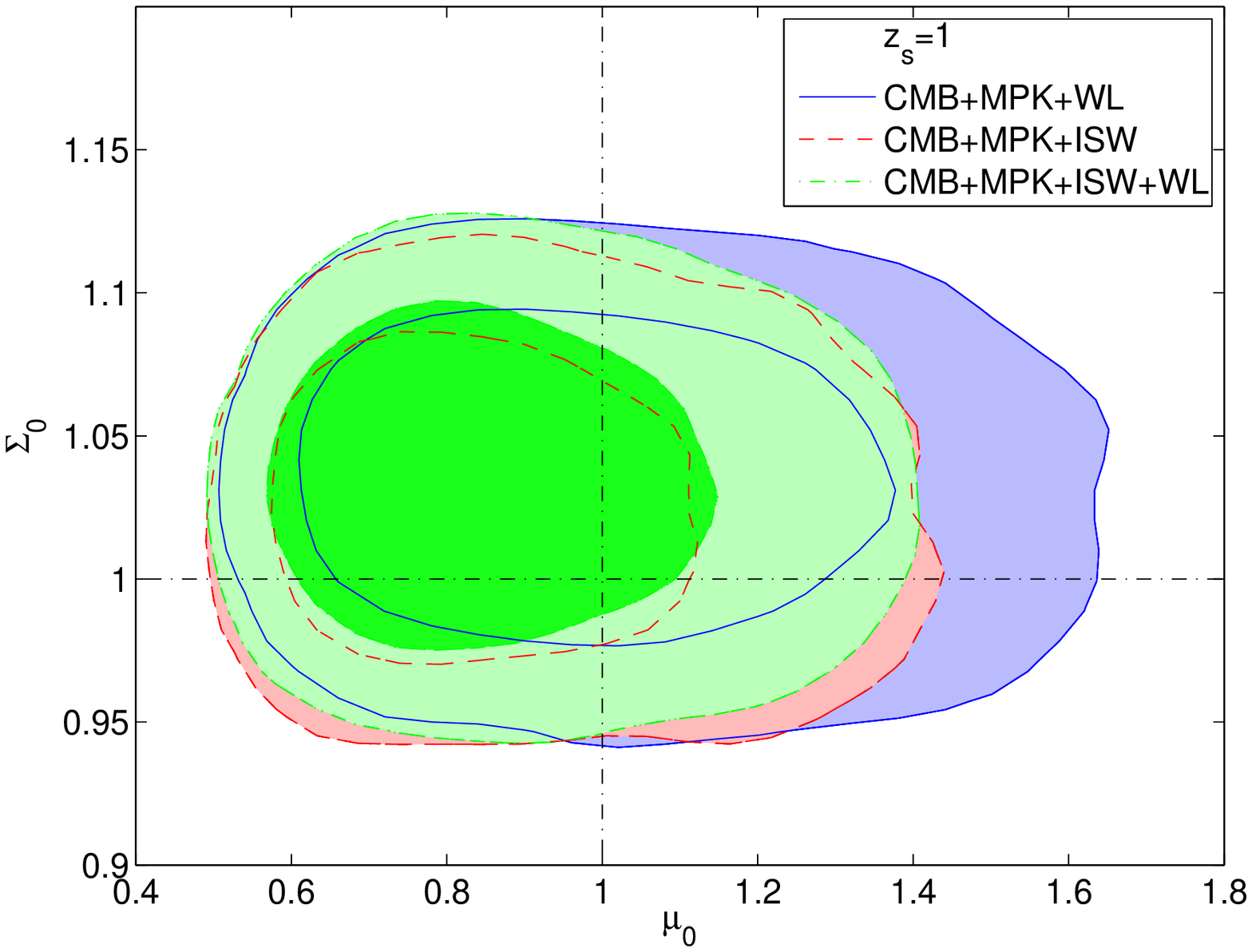}} &
{\includegraphics[width=2.in,height=2.in,angle=0]{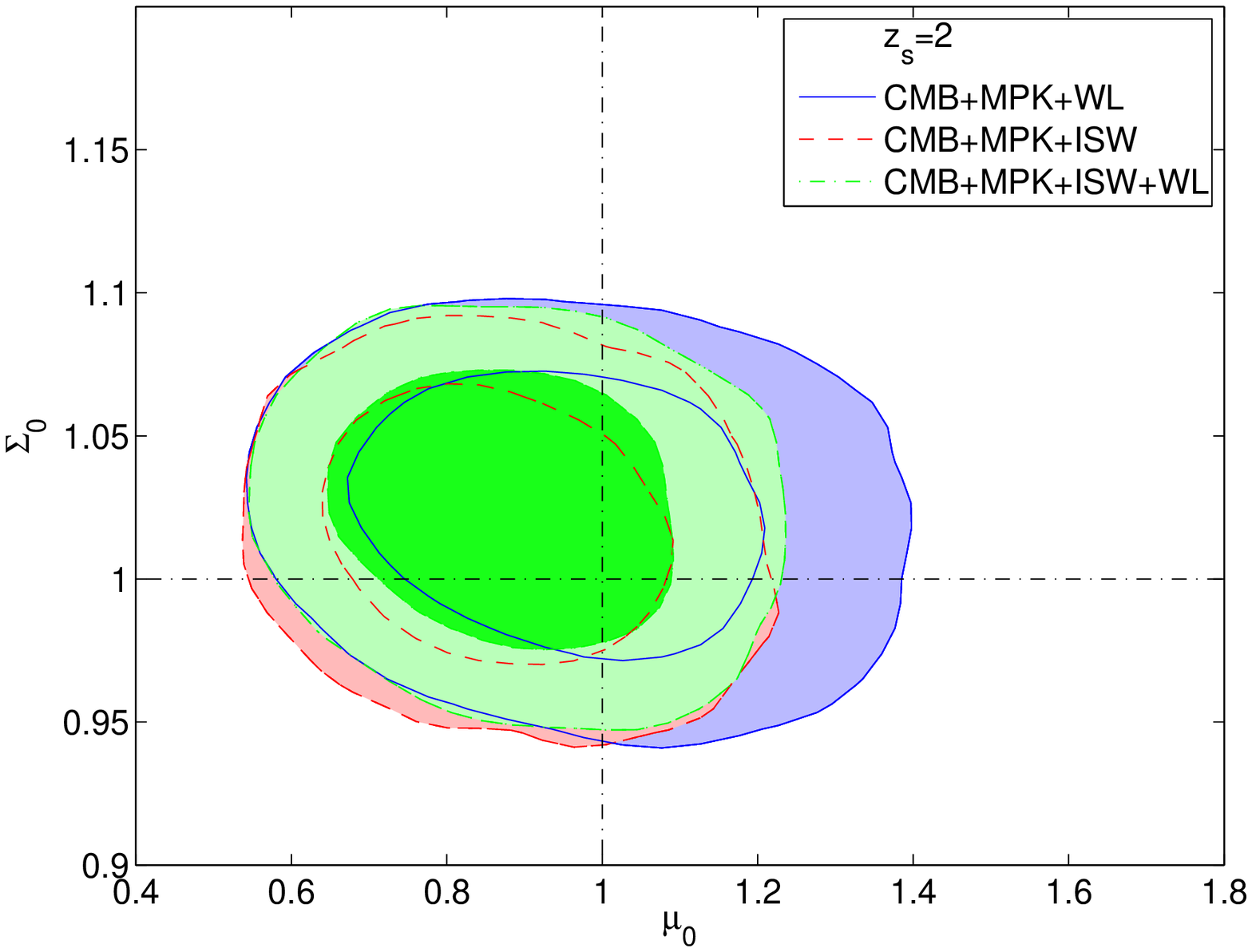}} &
{\includegraphics[width=2.in,height=2.in,angle=0]{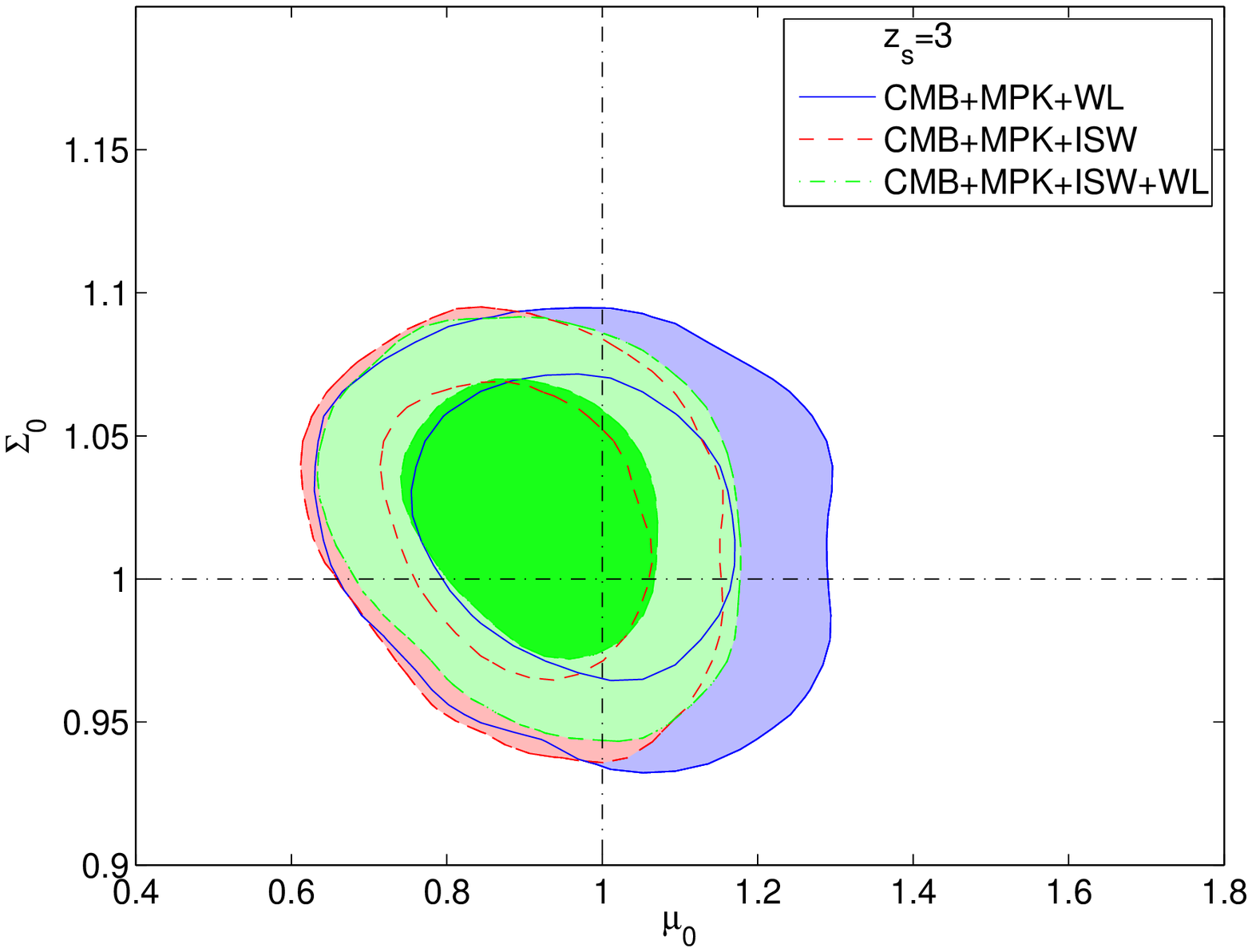}} \\
\hline  
\end{tabular}
\caption{\label{figure:obs2}
TOP-LEFT:$68\%$ and $95\%$ C.L. for the constraints on $\mu_0$ and $\Sigma_0$ of  {the second parametrization \cite{GongBo2010,MGCAMB}} with threshold redshift, $z_s=1$ and transition width fixed, $\Delta z=0.05$.  All contours are fit for CMB, SN, BAO, BBN, AGE and $H_0$. Contours enclosed by a solid line include WL, contours enclosed by a dashed line include WL and MPK, and contours enclosed by a dash-dotted line include WL, MPK, and ISW. TOP-CENTER: Same as LEFT, but for threshold redshift $z_s=2$.  TOP-RIGHT: Same as LEFT, but for threshold redshift $z_s=3$.
BOTTOM-LEFT:$68\%$ and $95\%$ C.L. for the constraints on $\mu_0$ and $\Sigma_0$ of  {the second parametrization \cite{GongBo2010,MGCAMB}} with threshold redshift, $z_s=1$ and transition width fixed, $\Delta z=0.05$.  All contours are fit for CMB, MPK, SN, BAO, BBN, AGE and $H_0$. Contours enclosed by a solid line include WL, contours enclosed by a dashed line include ISW, and contours enclosed by a dash-dotted line include WL and ISW.  BOTTOM-CENTER: Same as LEFT, but for threshold redshift $z_s=2$.  BOTTOM-RIGHT: Same as LEFT, but for threshold redshift $z_s=3$. The point $(1,1)$ indicates the GR values.} 
\end{center}
\end{figure}

Using the first parametrization, we find that the combination CMB+ISW+WL provides the strongest constraints on the parameters. We also observed that there seems some tension between current WL and MPK data. This reveals itself particularly well when comparing the FoM's of fits that include both these data sets to those that include only one of them eg. CMB+MPK+WL vs CMB+WL. Similarly, the combinations CMB+ISW+WL or CMB+MPK+ISW do better than the combination CMB+MPK+ISW+WL. 
In order to investigate this, we compared the best-fit models preferred by each combination of data sets finding that significant tensions exist between the best-fit values for the MG  parameters when using this parametrization. There are no significant tensions between the core cosmological parameters. 
This tension is illustrated in Fig. \ref{fig:MPKWLTENS} for both constraints on $Q_0$ and $R_0$ as well as the inferred parameters $\mathcal{G}_{}$ or $\Sigma_1$ and $\mathcal{V}_{}$ or $\mu_1$. 
 Also, in searching for the origin of this tension, we have both fixed the six core cosmological parameters to the WMAP7 values and additionally the time dependence parameter to $s=3$, yet it persists, significantly. We also find that the second parametrization below does not exhibit this particular tension between WL and MPK but shows other tension although substantially less pronounced.

\begin{figure}
\begin{center}
\begin{tabular}{|c|c|}
\hline

{\includegraphics[width=2.8in,height=2.in,angle=0]{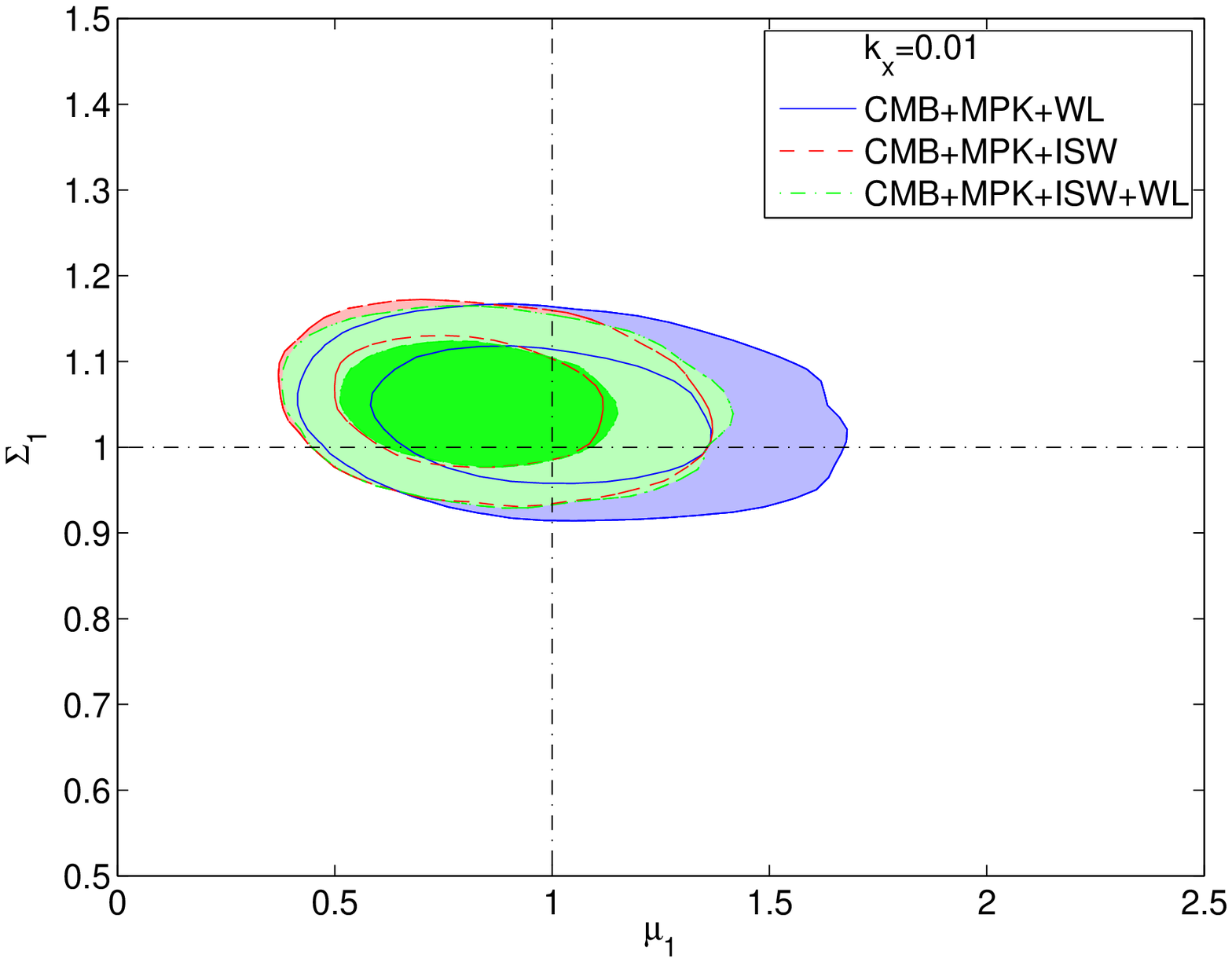}} &
{\includegraphics[width=2.8in,height=2.in,angle=0]{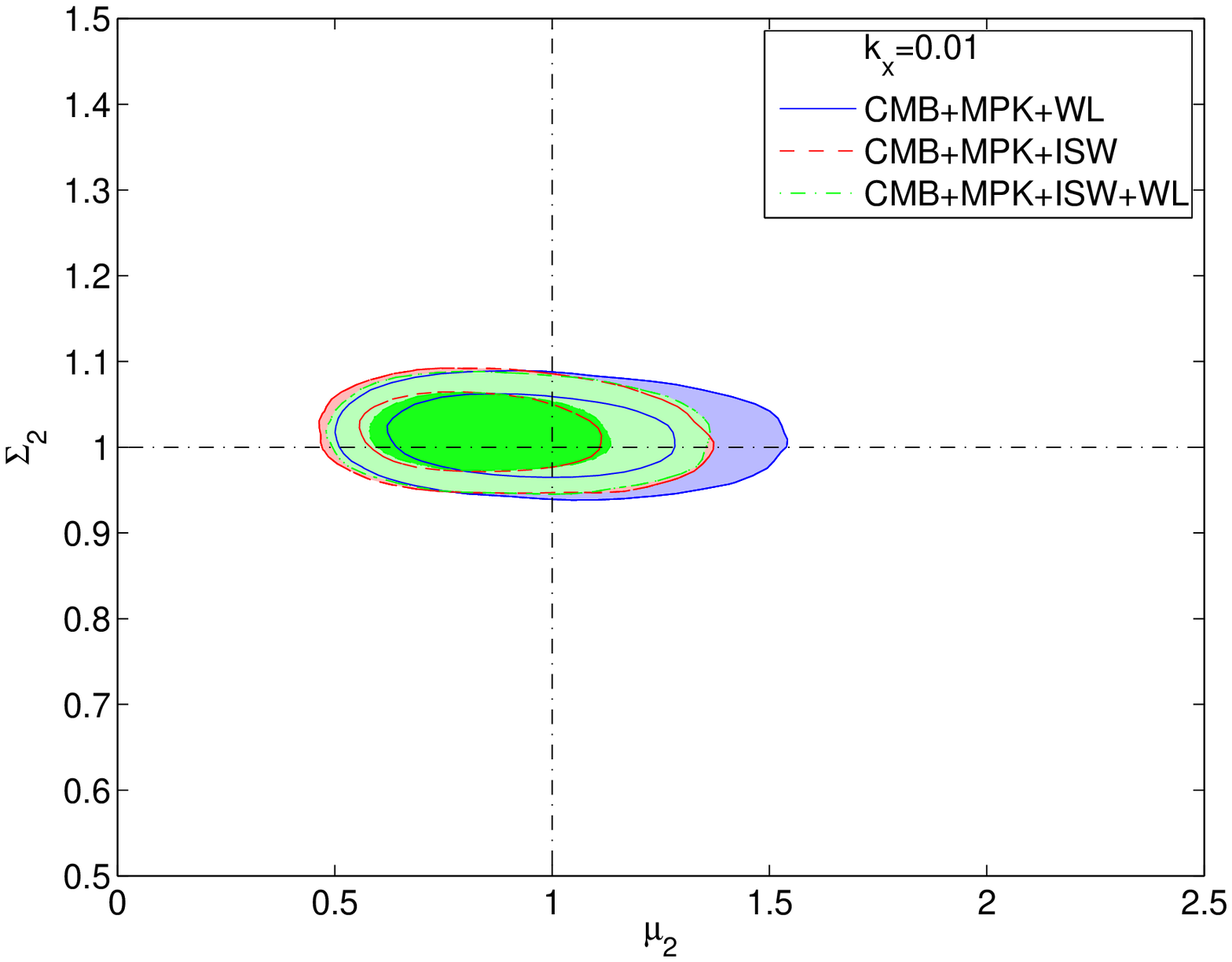}} \\
\hline  
{\includegraphics[width=2.8in,height=2.in,angle=0]{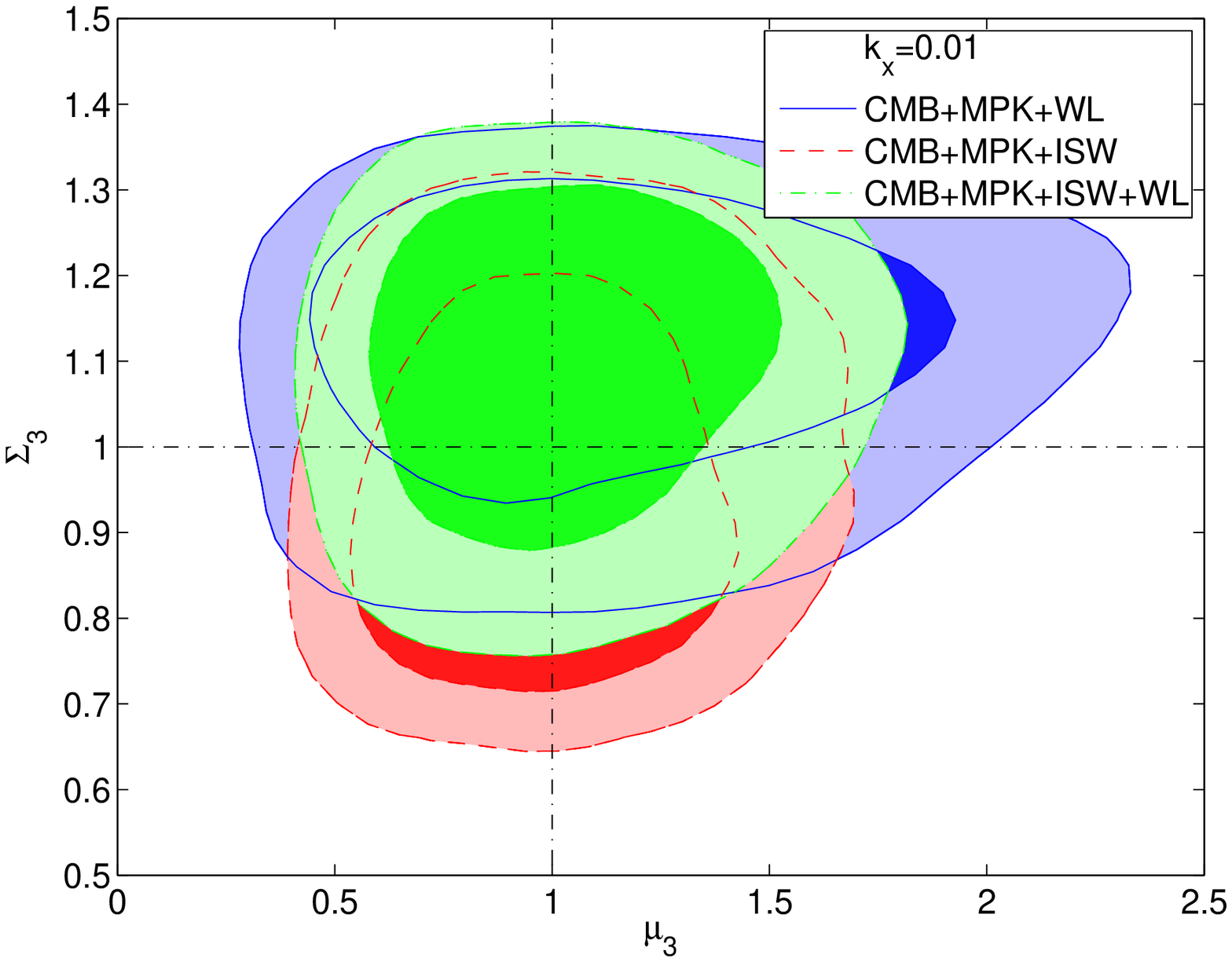}} &
{\includegraphics[width=2.8in,height=2.in,angle=0]{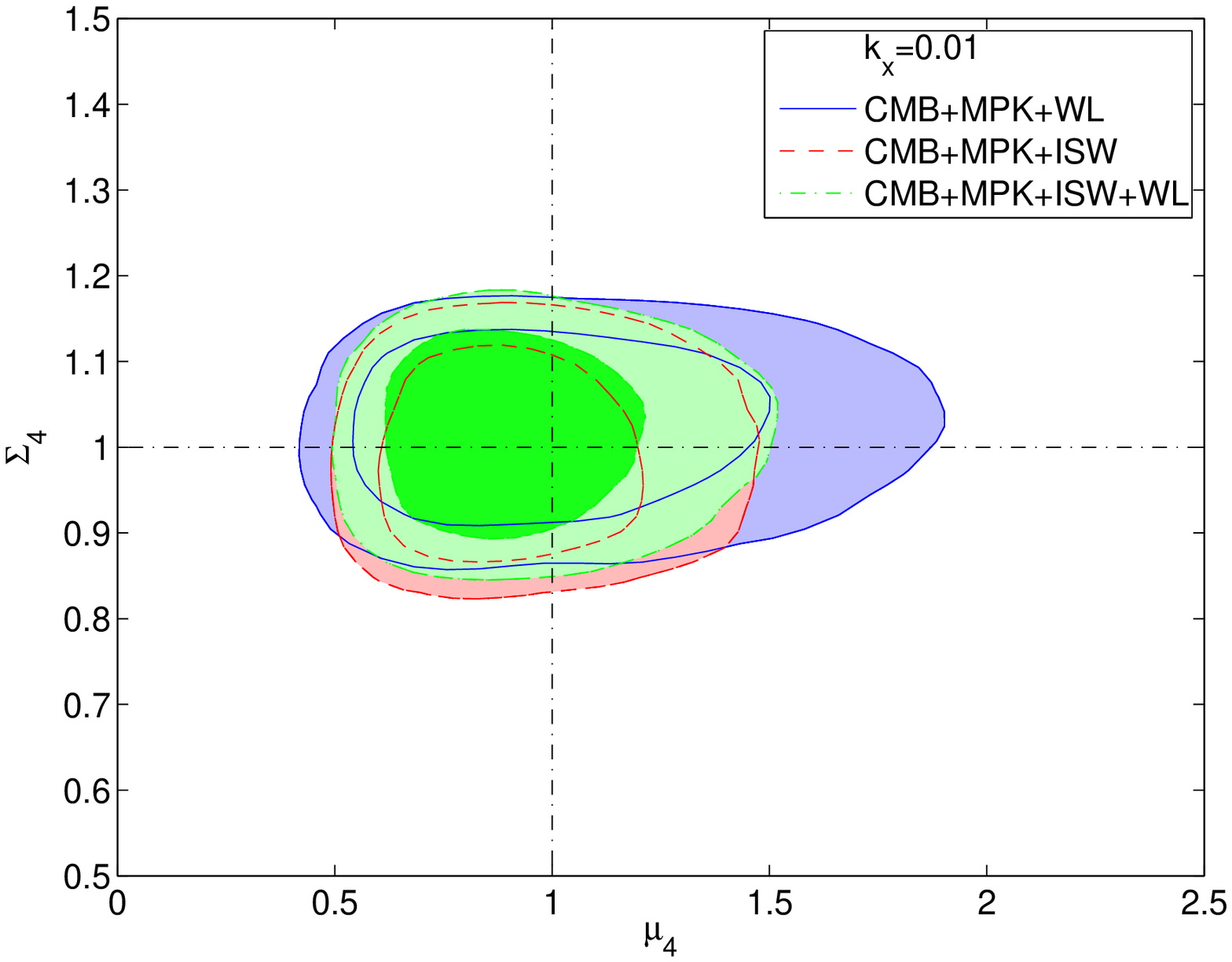}} \\
\hline
\end{tabular}
\caption{\label{figure:obs3}
$68\%$ and $95\%$ C.L. for the parameters $\mu_{i}$ and $\Sigma_{i}$ from  {the second parametrization \cite{GongBo2010,MGCAMB}} for redshift  and scale dependence as a binned $2 \times  2$ parametrization with $0<z\leq1,\,1<z\leq2$ and scale separation at $k_x=0.01\, h\, Mpc^{-1}$.   All contours are fit for CMB, MPK, SN, BAO, BBN, AGE and $H_0$.  Contours enclosed by a solid line include WL, contours enclosed by a dashed line include ISW, and contours enclosed by a dash-dotted line include WL and ISW. The point $(1,1)$ indicates the GR values.} 
\end{center}
\end{figure}

For the second parametrization, i.e. the one used in \texttt{MGCAMB} \cite{GongBo2010}, we derive uncertainties and FoM for $\mu(k,\,a)$ and $\Sigma(k,\,a)$ to avoid the strong degeneracy between $\mu(k,\,a)$ and $\eta(k,\,a)$. In order to allow for the parameters to be redshift (time) dependent and scale (wavenumber) dependent, we use a binned approach to look for deviation from GR, with no functional form assumed for the parameters.  First, however, we allow only redshift dependence  in order to compare to the first parametrization \cite{Bean2010} and look for deviation at late times, i.e. $z<z_s$ with a threshold redshift, $z_s = 0.76,\,1$, $2$, or $3$ and assume that at early times $\mu(k,\,a)=\eta(k,\,a) =1$.  As a remainder, the evolution in time is described by (\ref{eq:GBmuEvo}) and (\ref{eq:GBetaEvo}). We use the transition width in (\ref{eq:GBmuEvo}) and (\ref{eq:GBetaEvo}) to be both fixed,  $\Delta z=0.05$, so that the transition is almost a step function, as well as, allowed to vary in the interval $[0.01,\,0.5]$, becoming a nuisance parameter marginalized over. We allow these parameters to vary as follows: $-10<\mu_0<10$, $0<\Sigma_0<10$, and ruling out parameter combinations where $\eta_0<-1$. For each data set used, our results are given in Table \ref{table:obs} showing the FoMs and  $95\%$ confidence limits on  $\mu_0$ and $\Sigma_0$, as well as the inferred constraints on $\eta_0$, for this case (i.e. redshift (time) dependence only).  See Fig. \ref{figure:obs2} for comparisons of the $95\%$ confidence limit contours.  We find that the general relativity values, ($\mu_0=1$, $\Sigma_0=1$), are within $95\%$ confidence level for all data sets. The constraints from $z_s=2$ are tighter than $z_s=1$ because the earlier threshold redshift allows more change in gravitational potential and growth from the data sets, also noted in \cite{GongBo2010}. We also add constraints where we set the threshold redshift $z_s=3$ taking advantage of the higher redshift (i.e. $2<z\leq 3$) WL data in HST-COSMOS.  We know that the relevant redshift for cosmic acceleration is much smaller, but it is of interest to simply study the constraints obtained on the growth of structure using various available ranges of data.  From the physically motivated transition redshift (from deceleration to acceleration in an Lambda cold dark matter model) given by $1+z_{trans}=({2\Omega_{\Lambda}}/{\Omega_m})^{1/3}\approx 1.76$ (for example, see Eq. (14) in \cite{IshakReport} with $\Omega_{\Lambda}=0.73,\, \Omega_m=0.27$), we add $z_s=0.76$ for comparison, also.  The physically motivated transition redshift finds the weakest constraints and lowest FoM's due to less data below $z_s=0.76$, but all results are consistent with GR. 

The strongest constraints and highest FoM's we find come from using the combination CMB+MPK+ISW for both $z_s=1,\,2$, whereas, for $z_s=3$ the tightest constraint comes from using all the data sets including lensing, i.e. CMB+MPK+ISW+WL because with $z_s=3$, we allow more WL data below the threshold redshift to constrain the MG parameters which also yields a longer constant value at late times.  We find that the FoM for CMB+ISW is better than the one for CMB+MPK, which in turn is better than the FoM for CMB+WL. This shows that the current ISW data provides indeed the strongest constraints on the MG parameters. 
Also, as shown in Table \ref{table:obs}, the allowance of the transition width, $\Delta z$, to vary (i.e. $[0.01,\,0.5]$) loosens the constraints and reduced the FoM for all $z_s$ models.  
Now, as found for the first parametrization, here also, there is a slight decrease in the FoM and weakening of the constraints (for $z_s = 1,\,2$) when WL is added to CMB+MPK+ISW, again, due to some tensions between the preferred MG  parameters by different data sets.  To investigate these tensions, here we also fixed the core  cosmological parameters and left the MG  parameters to vary. As a result, the FoM's increase consistently (but only moderately in some cases) when adding a data set, indicating that some of the tension was reduced but not removed. In fact, a closer look at the best-fit MG  parameters preferred by different data sets shows that, while reduced, a non-negligible scatter is still present. This explains the only moderate improvements obtained in some cases. This indicates that there are non-negligible tensions intrinsic to the MG parameters preferred by different data sets.}

\begin{figure}
\begin{center}
\begin{tabular}{|c|c|}
\hline

{\includegraphics[width=2.8in,height=2.in,angle=0]{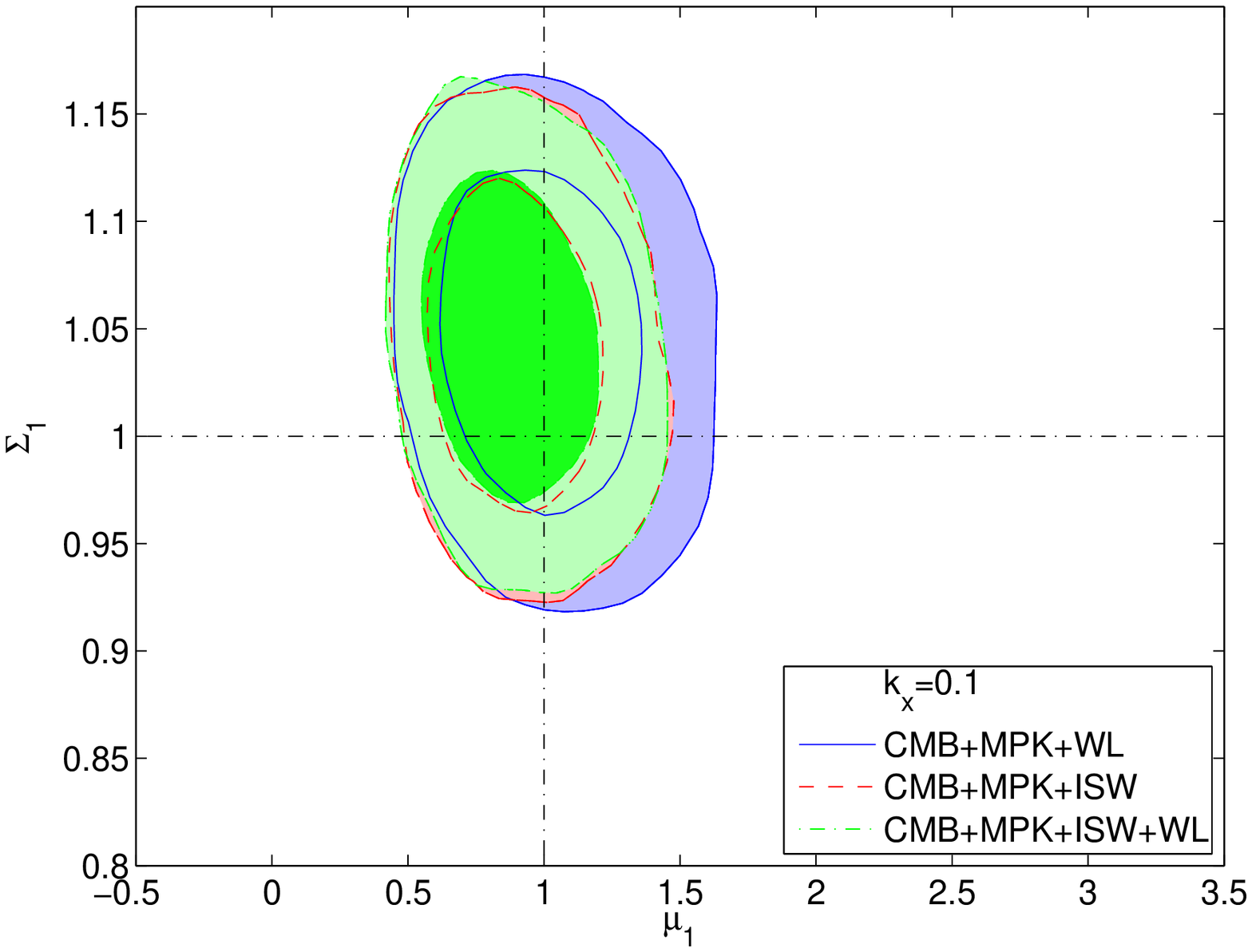}} &
{\includegraphics[width=2.8in,height=2.in,angle=0]{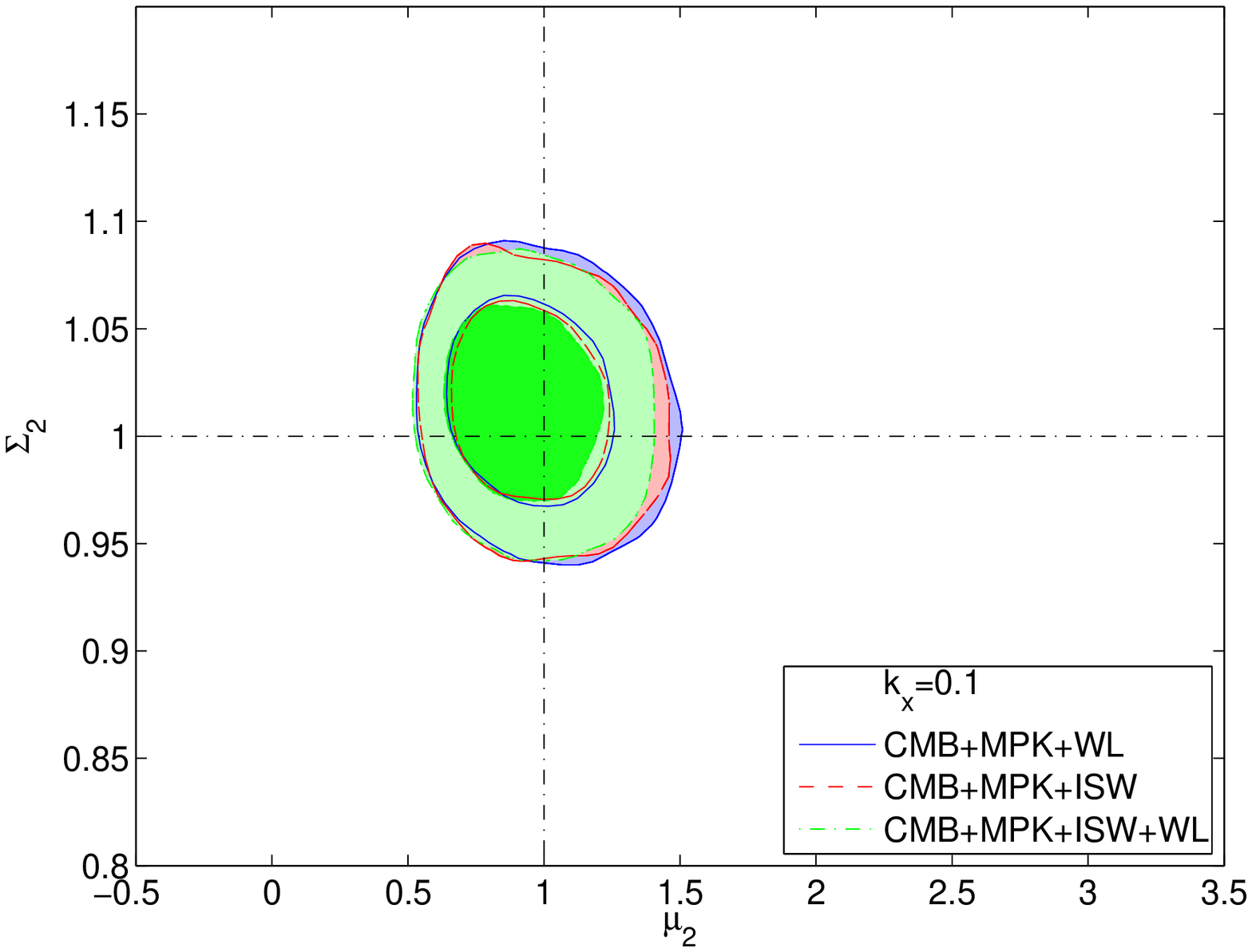}} \\
\hline  
{\includegraphics[width=2.8in,height=2.in,angle=0]{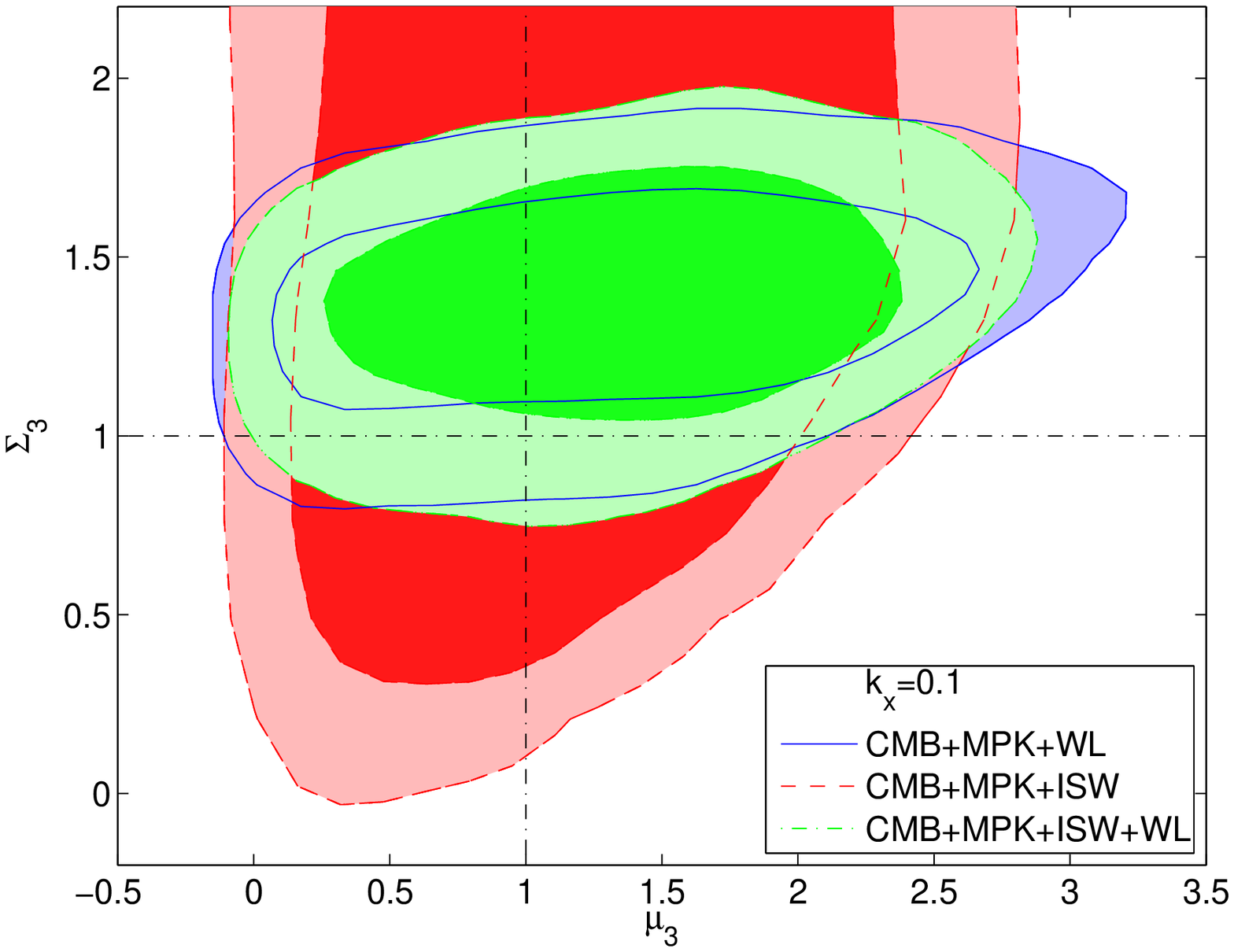}} &
{\includegraphics[width=2.8in,height=2.in,angle=0]{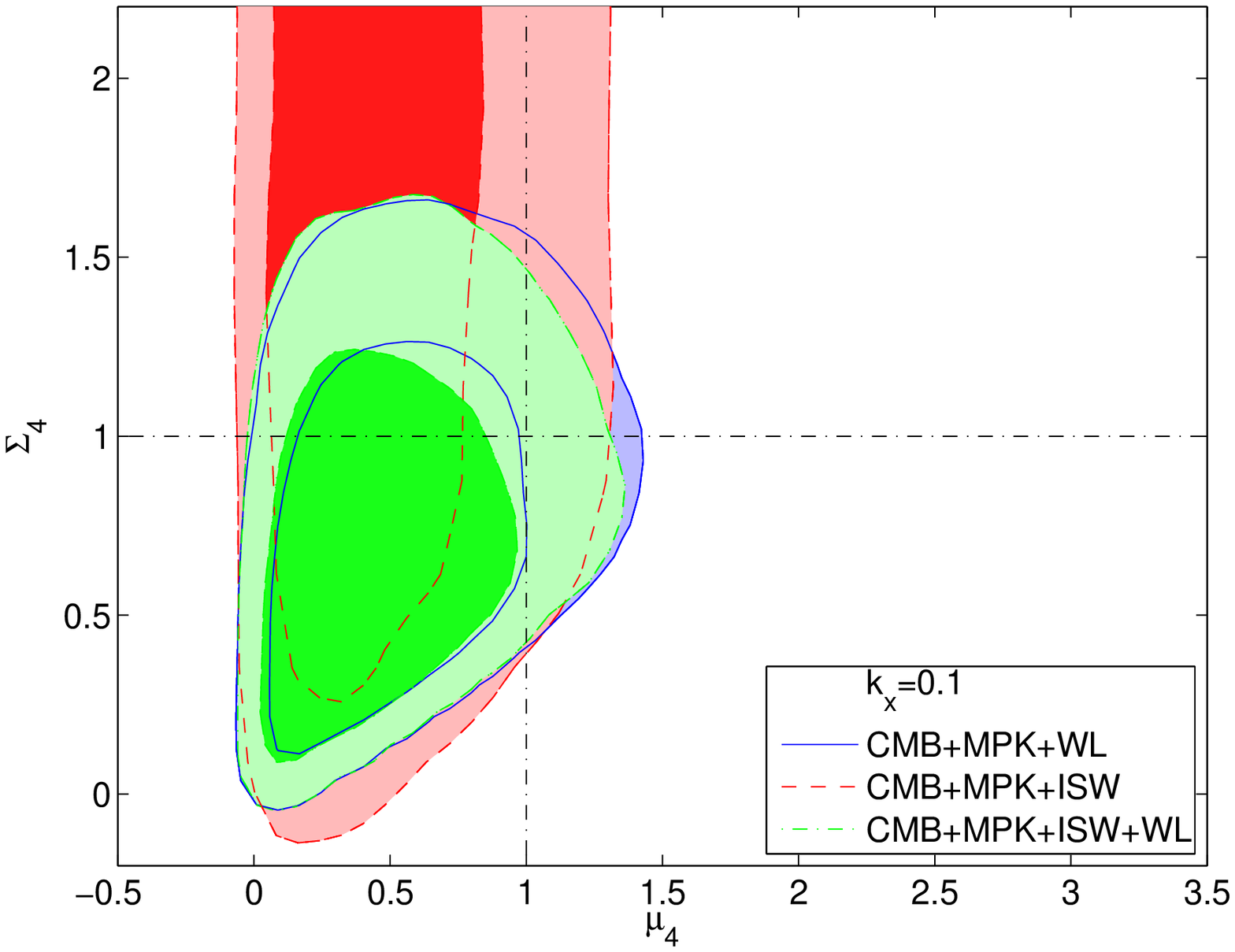}} \\
\hline
\end{tabular}
\caption{\label{figure:obs4}
$68\%$ and $95\%$ C.L. for the parameters $\mu_{i}$ and $\Sigma_{i}$ from  {the second parametrization \cite{GongBo2010,MGCAMB}} for redshift  and scale dependence as a binned $2 \times  2$ parametrization with $0<z\leq 1,\,1<z\leq 2$ and scale separation at $k_x=0.1\, h\, Mpc^{-1}$.   All contours are fit for CMB, MPK, SN, BAO, BBN, AGE and $H_0$.   Contours enclosed by a solid line include WL, contours enclosed by a dashed line include ISW, and contours enclosed by a dash-dotted line include WL and ISW. The point $(1,1)$ indicates the GR values.} 
\end{center}
\end{figure}

Next, we calculate constraints and FoM's while we allow for simultaneous scale (wavenumber)  and redshift (time) dependence by using variants of a  $2 \times 2$ binned approach (called pixilation in \cite{GongBo2010}), see earlier discussion and Table \ref{table:Grid} for details.  We alternate between ISW and WL and their combination, while using CMB, MPK, BAO, SN, BBN, AGE and $H_0$ in all cases. We provide the FoM's and $95\%$ confidence limits on $\mu_{i}$ and $\Sigma_{i}$, as well as, the inferred ones on $\eta_{i}$ in Table \ref{table:obs2} for scale-dependence separators at $k_x= 0.01\, h\, Mpc^{-1}$ and $k_x= 0.1\, h\, Mpc^{-1}$, see Figs. \ref{figure:obs3} and \ref{figure:obs4} for the comparisons in their C.L. contours.  General relativity values for $\mu_{i}=1$ and $\Sigma_{i}=1$ are within all constraints for $\mu_{i}$ and $\Sigma_{i}$ for all the combination of data sets. Unlike the work that uses a binned parametrization and CFHTLS \cite{GongBo2010, DL2010}, we do not find a deviation from GR with $\Sigma_3$. But as expressed there as well, this is due to the fact that we are using the refined HST-COSMOS data by \cite{Schrabback2010} which does not have the known systematic effect (i.e. residual systematic 'bump') as discussed in the papers using CFHTLS, \cite{GongBo2010, DL2010, FuCFHTLS,CFHTNote}.

 \begin{center}
\begin{table}[t]
\begin{tabular}{|c|c|c|c|c|c|c|}\hline
\multicolumn{7}{|c|}{\bfseries Constraints for  $\{\mu_{i},\,\Sigma_{i}\}$ binned parametrization $0<z\leq1,\,1<z\leq2$}\\ \hline
 &\multicolumn{3}{|c|}{$k_x=0.01$} & \multicolumn{3}{|c|}{$k_x=0.1$}\\ \hline 
  data sets &WL &ISW&ISW, WL&WL &ISW&ISW, WL\\ \hline
  FoM$_{1}$ &  75.92 & 101.7 & 101.2 &  80.63 & 99.47 & 97.62\\ \hline 
$\mu_{1}$ & $[ 0.559,\,1.567]$ &$[ 0.480,\,1.227]$ & $[ 0.494,\,1.309]$& $[ 0.587,\,1.540]$ &$[ 0.542,\,1.356]$ & $[ 0.535,\,1.357]$\\ \hline
$\Sigma_{1}$ &  $[ 0.940,\,1.138]$ &$[ 0.957,\,1.145]$ & $[ 0.955,\,1.139]$ &  $[ 0.946,\,1.141]$ &$[ 0.949,\,1.136]$ & $[ 0.951,\,1.139]$\\ \hline 
$\eta_{1}$ &  $[ 0.281,\,2.815]$ &$[ 0.605,\,3.486]$ & $[ 0.563,\,3.347]$ &  $[ 0.322,\,2.644]$ &$[ 0.493,\,2.939]$ & $[ 0.496,\,3.004]$\\ 
 (derived)&&&&&&\\ \hline 
FoM$_{2}$ & 148.9 & 176.3 & 184.2 & 161.7 &172.4 & 180.5 \\ \hline 
$\mu_{2}$ & $[ 0.614,\,1.456]$ &$[ 0.563,\,1.289]$ & $[ 0.580,\,1.294]$ & $[ 0.633,\,1.413]$ &$[ 0.635,\,1.382]$ & $[ 0.616,\,1.335]$\\ \hline 
$\Sigma_{2}$ &  $[ 0.955,\,1.072]$ &$[ 0.959,\,1.075]$ & $[ 0.959,\,1.072]$ &  $[ 0.955,\,1.075]$ &$[ 0.956,\,1.072]$ & $[ 0.957,\,1.072]$\\ \hline 
$\eta_{2}$ &  $[ 0.376,\,2.332]$ &$[ 0.546,\,2.649]$ & $[ 0.554,\,2.526]$&  $[ 0.416,\,2.238]$ &$[ 0.452,\,2.229]$ & $[ 0.503,\,2.329]$\\
 (derived)&&&&&&\\ \hline
FoM$_{3}$ &  18.95 & 22.94 & 25.61 &  5.989 &1.098 & 6.755\\ \hline 
$\mu_{3}$ & $[ 0.439,\,2.073]$ &$[ 0.509,\,1.588]$ & $[ 0.544,\,1.663]$ & $[ 0.093,\,2.771]$ &$[ 0.122,\,2.654]$ &$[ 0.159,\,2.519]$  \\ \hline 
$\Sigma_{3}$ &   $[ 0.851,\,1.321]$ &$[ 0.708,\,1.268]$ & $[ 0.818,\,1.319]$ &   $[ 0.903,\,1.824]$ &$[ 0.546,\,5.391]$ & $[ 0.871,\,1.825]$\\ \hline 
$\eta_{3}$&  $[ 0.069,\,4.035]$ &$[ 0.129,\,2.889]$ & $[ 0.244,\,3.066]$&  $[ 0.052,\,26.31]$ &$[ 0.051,\,35.69]$ & $[ 0.037,\,14.78]$\\
 (derived)&&&&&&\\ \hline 
FoM$_{4}$ & 43.94 & 62.03 & 63.93 &  8.878 & 2.014 & 9.214\\ \hline 
$\mu_{4}$ &  $[ 0.553,\,1.754]$ &$[ 0.588,\,1.404]$ & $[ 0.593,\,1.427]$ &  $[ 0.061,\,1.269]$ &$[ 0.032,\,1.288]$ & $[ 0.052,\,1.194]$ \\ \hline 
$\Sigma_{4}$ &  $[ 0.887,\,1.148]$ &$[ 0.858,\,1.134]$ & $[ 0.878,\,1.144]$  &  $[ 0.173,\,1.552]$ &$[ 0.381,\,5.181]$ & $[ 0.167,\,1.563]$ \\ \hline 
$\eta_{4}$ &  $[ 0.152,\,2.708]$ &$[ 0.376,\,2.424]$ & $[ 0.398,\,2.458]$&  $[ 0.052,\,18.46]$ &$[ 0.024,\,169.9]$ & $[ 0.031,\,22.27]$\\  
(derived)&&&&&&\\ \hline  \hline

\hline
\end{tabular}
\caption{\label{table:obs2}
$95\%$ C.L. for the parameters $\mu_{i}$, $\eta_{i}$, and $\Sigma_{i}$ from  {the second parametrization \cite{GongBo2010,MGCAMB}} for redshift  and scale dependence as a binned $2 \times  2$ parametrization with $0<z\leq1,\,1<z\leq2$ and scale separation at $k_x=0.01\, h\, Mpc^{-1}$ and $k_x=0.1\, h\, Mpc^{-1}$ .  The data used is described in the text above and indicated as: ISW-galaxy cross correlations (ISW), weak-lensing tomography (WL). In all cases, the data used is combined with WMAP7 temperature and polarization spectrum (CMB), the matter power spectrum (MPK), SN, BAO, BBN, AGE and $H_0$.  The FoM for each data set and $\{\mu_{i},\,\Sigma_{i}\}$ bin is also given.
}
\end{table}
\end{center}

 Another observation that can be made is that with the data sets used in this paper, the binning with scale separator at $k_x=0.01\, h\, Mpc^{-1}$ better constrains $\{\mu_{i},\,\Sigma_{i}\}$ by the balancing of the number of data points within the bins, so within each bin, we are taking maximum advantage of the available data.  Whereas, the wave number separator at $k_x=0.1\, h\, Mpc^{-1}$ does not equally balance the available data, i.e. $\{\mu_{3},\,\Sigma_{3}\}$ and $\{\mu_{4},\,\Sigma_{4}\}$ are not well constrained by ISW because most ISW data points are below $k_x=0.1\, h\, Mpc^{-1}$, which allows WL to dominate the $\{\mu_{3},\,\Sigma_{3}\}$ and $\{\mu_{4},\,\Sigma_{4}\}$ bins.  However, for bins with a scale separator at $k_x= 0.01\, h\, Mpc^{-1}$, the FoM increases for  $\{\mu_{i},\,\Sigma_{i}\}$ in each bin beginning with the lowest for WL, then ISW, and the highest for the combination of ISW and WL because of the balance of data points, see Table \ref{table:obs2}, see Fig. \ref{figure:obs3} (again CMB, MPK, BAO, SN, BBN, AGE and $H_0$ are included each time). There is a slight decrease of the FoM for the $\{\mu_{1},\,\Sigma_{1}\}$ bin because there is not enough WL data in this bin to increase the FoM when added to ISW. Some tension between ISW and WL data sets is more explicitly seen in bins $\{\mu_{1},\,\Sigma_{1}\}$ and $\{\mu_{2},\,\Sigma_{2}\}$ (see Table \ref{table:obs4}) for both scale separators. Future surveys and other data sets may work better with scale separator at $k_x=0.1\, h\, Mpc^{-1}$ \cite{Pogosian2010}. 

\begin{figure}
\begin{center}
\begin{tabular}{|c|c|}
\hline

{\includegraphics[width=2.8in,height=2.in,angle=0]{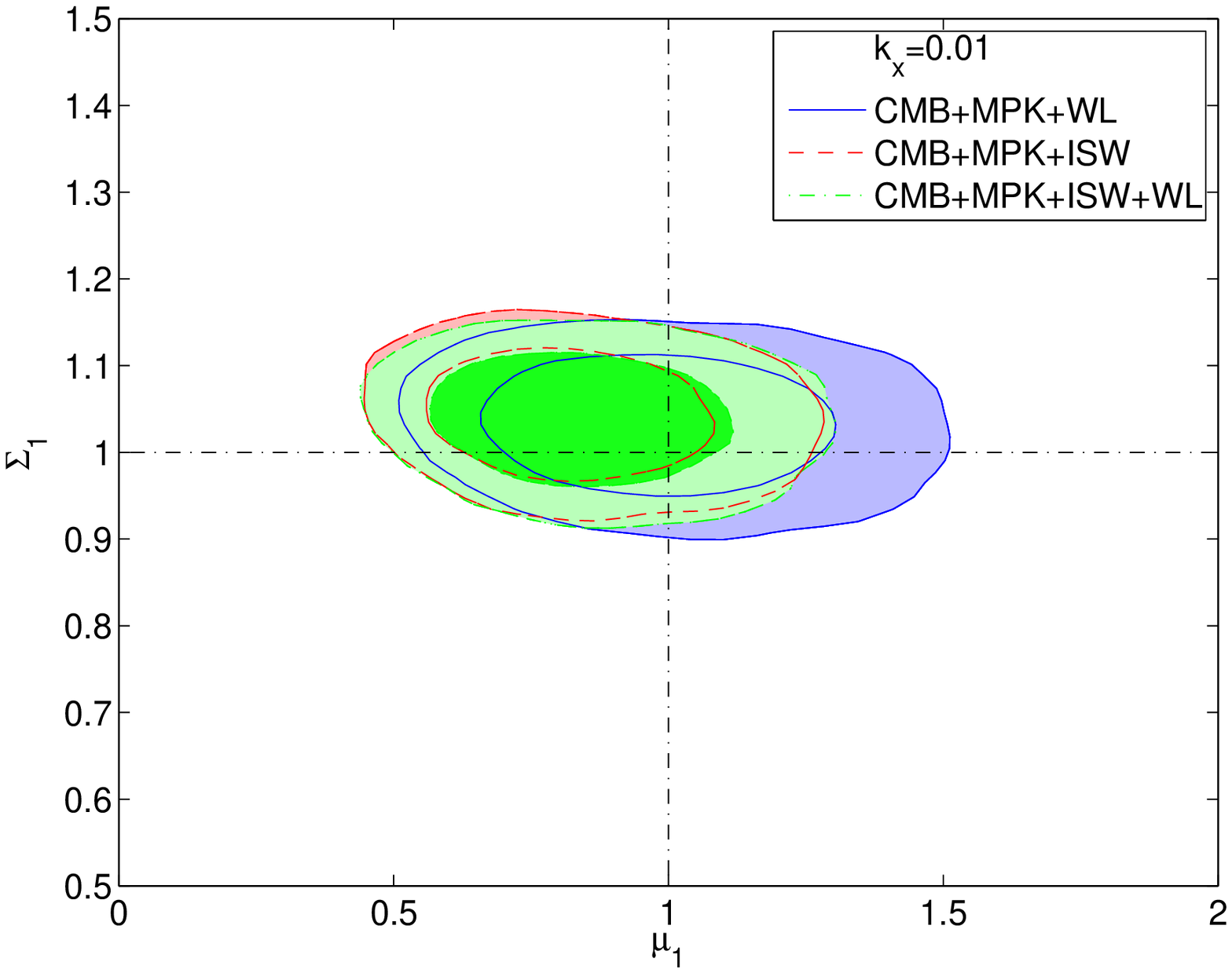}} &
{\includegraphics[width=2.8in,height=2.in,angle=0]{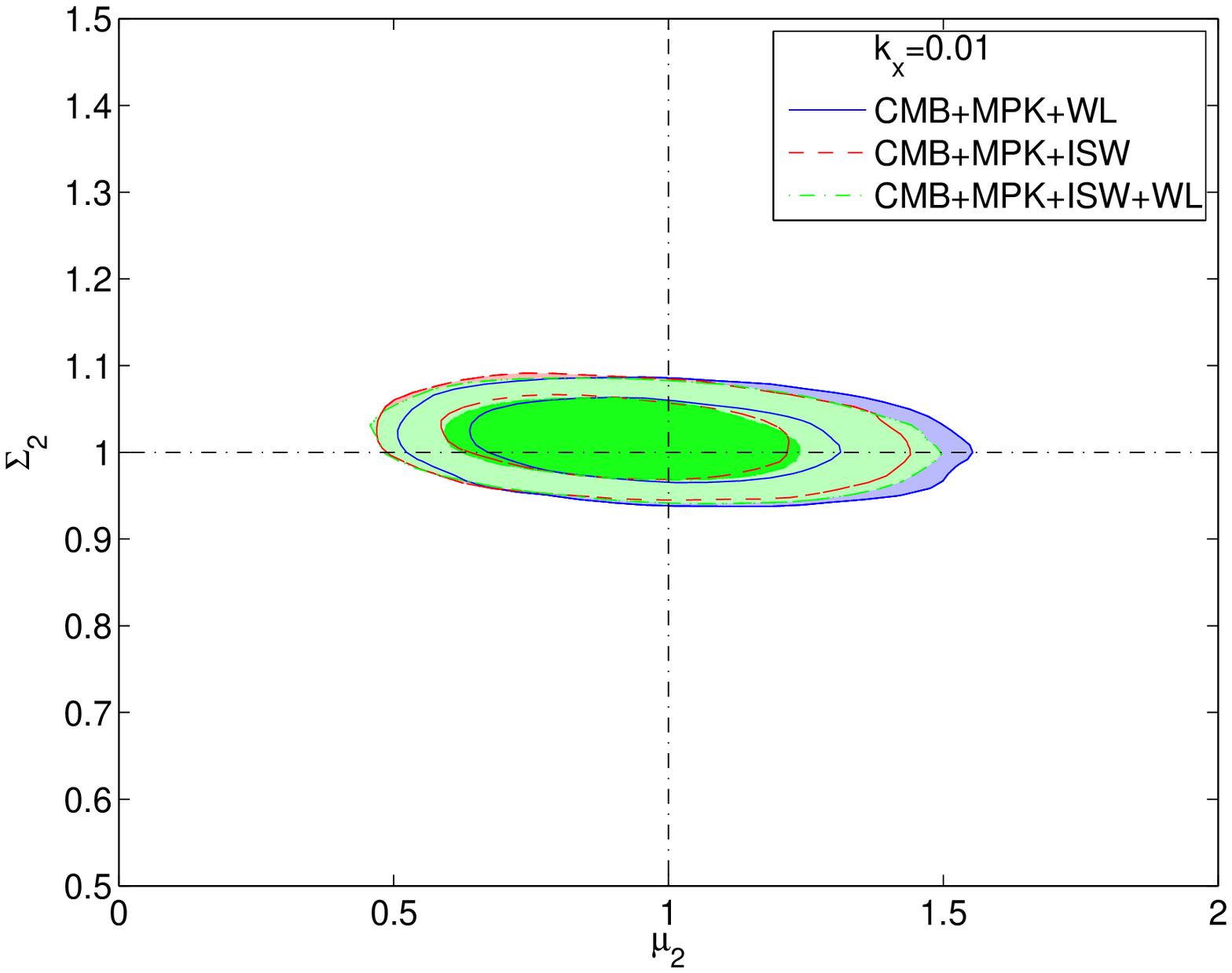}} \\
\hline  
{\includegraphics[width=2.8in,height=2.in,angle=0]{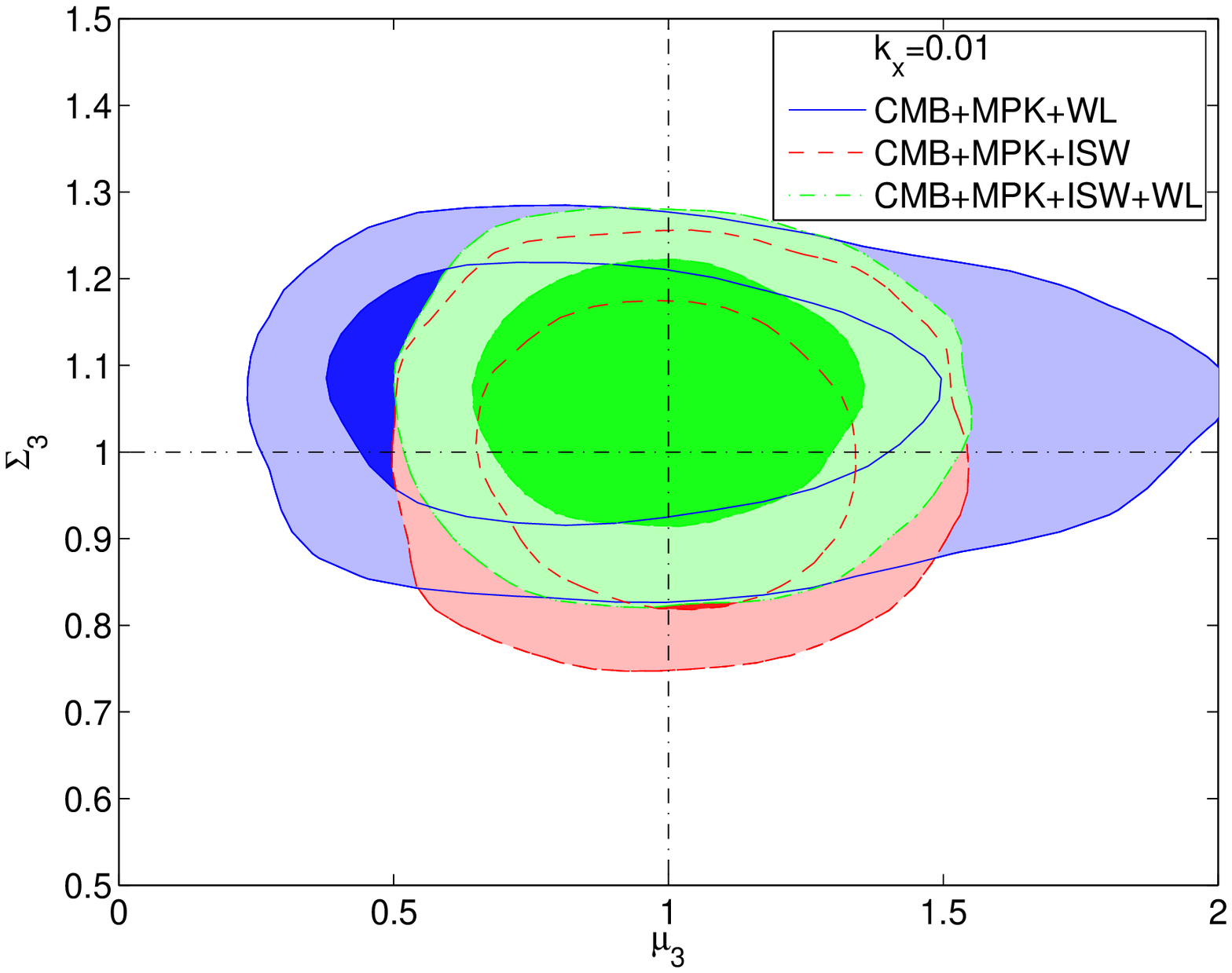}} &
{\includegraphics[width=2.8in,height=2.in,angle=0]{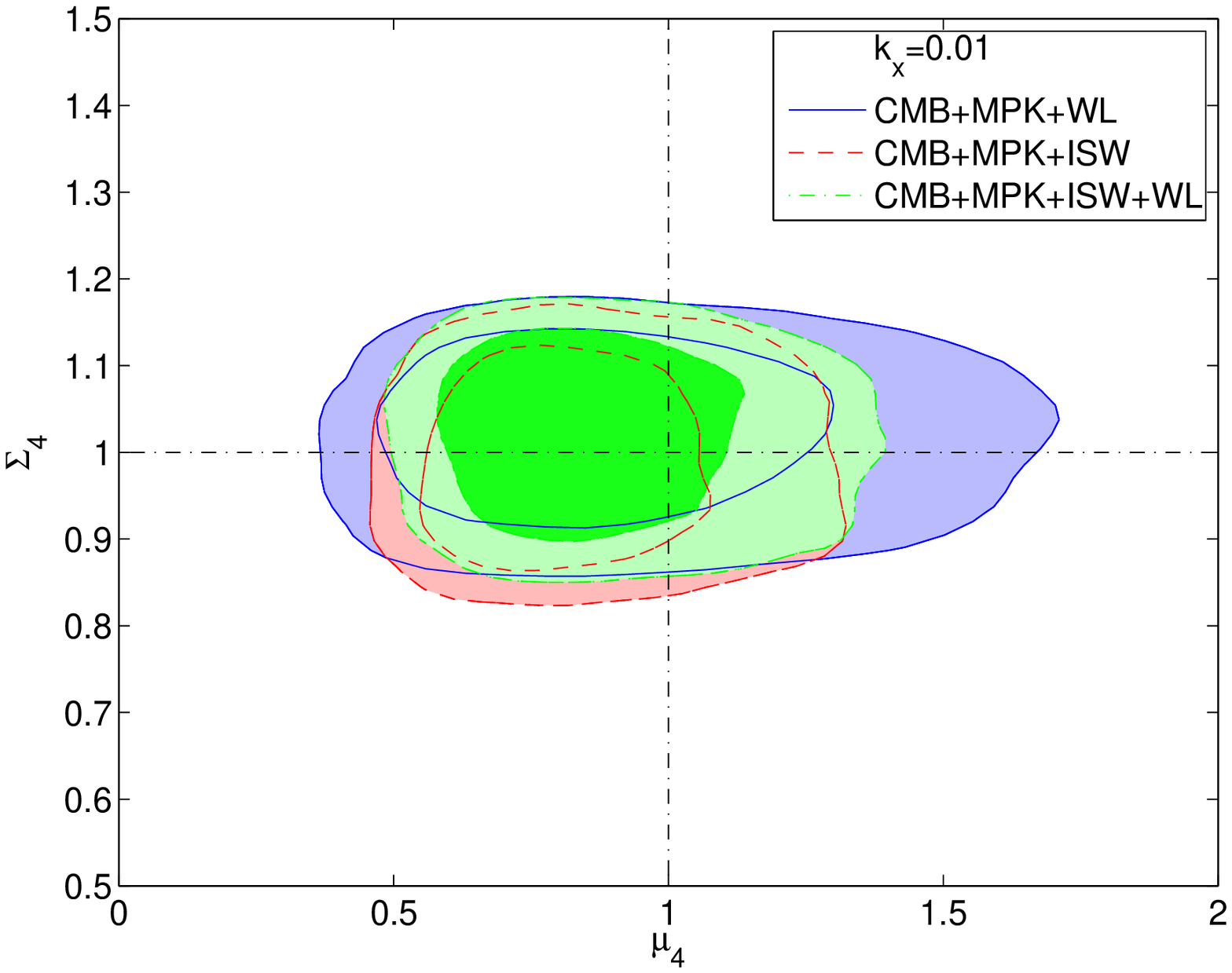}} \\
\hline
\end{tabular}
\caption{\label{figure:obs5}
$68\%$ and $95\%$ C.L. for the parameters $\mu_{i}$ and $\Sigma_{i}$ from  {the second parametrization \cite{GongBo2010,MGCAMB}} for redshift  and scale dependence as a binned $2 \times  2$ parametrization with $0<z\leq 1.5,\,1.5<z\leq 3$ and scale separation at $k_x=0.01\, h\, Mpc^{-1}$.   All contours are fit for CMB, MPK, SN, BAO, BBN, AGE and $H_0$.  Contours enclosed by a solid line include WL, contours enclosed by a dashed line include ISW, and contours enclosed by a dash-dotted line include WL and ISW. The point $(1,1)$ indicates the GR values.} 
\end{center}
\end{figure}

We give the results for the third binned method in Table \ref{table:obs4} and show the relevant ellipses for the FoM's in Fig. \ref{figure:obs5}.  The constraints in Table \ref{table:obs4} are tighter than the other binning results in Table \ref{table:obs2} and have higher FoM's corresponding to smaller ellipses in Fig. \ref{figure:obs5}.  The improvements in the constraints and FoM's comes from larger redshift bins allowing more data to constrain the parameters (called an \textit{accumulation effect} by \cite{GongBo2010}) and these larger redshift bins also allowing a longer constant value for late times.  The $\{\mu_{2},\,\Sigma_{2}\}$ bin is slighter looser in this third binned method because more of the ISW data points are in the $\{\mu_{1},\,\Sigma_{1}\}$ bin, now, due to a larger redshift bin, i.e. $0<z\leq 1.5$, than in the previous two binned methods.  We still observe some small tension when adding WL to the ISW data set combination with the FoM's and contour plots, and it is more easily seen for example in the $\{\mu_{1},\,\Sigma_{1}\}$ and $\{\mu_{2},\,\Sigma_{2}\}$ bins in Table \ref{table:obs4} where the  {FoM's} rather decrease.

Before summarizing our results further below it is worth discussing the possible physical origins of the tensions on best-fit modified gravity parameters prefered by different combinations of data sets. To further explore this we decided to fix the core cosmological parameters to their WMAP7 values and use each set individually and removing the CMB data set since its data points will most likely dominate parameter constraints, and likely wash out some of the tension in our best-fit parameter spaces. Our results expectedly gave us weak constraints and for both the first parametrization and the second parametrization (in its binned form), we continued to observe tensions where they had previously existed (the tension in the unbinned parametrization was removed by fixing the core cosmological parameters). The best-fit parameter spaces obtained did give us further insight into the preferred parameter values of each data set. For the first parametrization we saw that the best-fit for MPK preferred the lowest value of $s$ out of all the best-fits.  Additionaly, in contrast to WL best-fits, it had negative best-fit values for $R$ and $\mathcal{V}$, though the best-fit $Q$ and $D$ parameters are actually not that different between.  Combining the two data sets however we find that a best-fit for $s$ that is higher than either of the two data sets individually, additionally a higher $Q$ and correspondingly more negative $\mathcal{V}$ and $R$ are preferred. For the binned second parametrization we looked for tension between the MPK+ISW and WL data sets. In the first bin, the WL data set had a much higher best-fit $\mu$ and $\Sigma$ values than the MPK+ISW data set.  The MPK+ISW+WL data set though has the lowest best-fit $\Sigma_1$ and a similar besti-fit $\mu_1$ to the MPK+ISW data set again showing the tension when combining these data sets. These markedly different best-fit values in the modified gravity parameter space between the individual and combined data sets highlight the tension but do not explain it. 

Working again with the first parametrization to explore the tension there, we next looked at the ouptut matter power spectrum for each of the best-fit parameter sets. This showed just how much an exacerbation the parameter $s$ may be causing. The amplitude of the matter power spectrum produced with the best-fit WL parameter set was 2 orders of magnitude greater than the amplitude of the spectrum produced by the best-fit MPK parameter set.  Upon seeing this marked difference in amplitudes one can recall that in calculating the likelihood for MPK the amplitude of the power-spectrum is treated as a nuisance parameter \cite{BAOReid}.  This may be one reason for the tension when combining data sets, as a number of MG parameter combinations can produce matter-power spectra with similar shapes in the k-range where SDSS-DR7 data is present but with different amplitudes. The matter power spectrum is of course influenced by the growth of structure within a given gravity model. 

We recall again from above that the ISW-effect is dependent on the quantity
$(\dot\phi+\dot\psi)^2$, while the weak-lensing signal is dependent on the
quantity $\phi+\psi$. Furthermore, there has been much discussion in previous works \cite{GongBo2010,Bean2010,Daniel2010} as to how different values of the various MG parameters in the various parametrizations affect cosmological observations. We will recall some of that discussion related to the growth of structure in order to examine how different MG parameters will affect the matter-power spectrum. In the matter dominated era, the growth of structure (matter overdensities) on subhorizon scales (but still within the linear regime) is governed by the usual equation $\ddot\Delta_m + \mathcal{H}\dot\Delta_m +k^2\psi = 0$. Superhorizon growth of matter overdensities is a bit more complicated. We refer the reader to Eq. (A5) of \cite{Daniel2010} for a full expression of the growth equation in the matter dominated era ($\mu$ there should not be confused with the $\mu$ in this paper as they are different). As mentioned in the previous reference a $k^2\psi$ term will be present at all scales (it is actually scale-independent through Eq. (\ref{eq:GBmu})). This term  is affected directly by the parameters $\mu$ or $\mathcal{V}$ from the second and third parametrizations respectively, and correspondingly the product of $Q$ and $R$ from the first parametrization. An increase (decrease) in these quantites from their GR values will increase (decrease) the magnitude of the gravitational potential $\psi$. This increase (decrease) in the magnitude of the potential in turn combats (enhances) the suppression of the growth caused by the accelerated expansion of the universe, therebye enhancing (suppressing) the growth relative to its GR value on all scales.  This will in turn increase (decrease) the overall amplitude of the matter power spectrum and the weak-lensing correlations relative to their GR values. Though not explicitly shown in the aforementioned equation for superhorizon growth, the next dominant term in the growth equation on those scales involves the derivatives of the potentials. Thus they will involve the derivatives of the MG parameters (these are shown in the mentioned equation though). In the second parametrization the derivatives do not matter much as they are zero for most of the evolution. For the first parametrization, however, the derivatives directly involve the parameter $s$ as a multiplicative factor so the parameter $s$ affects the growth on these scales in more ways than just govering how long the gravity modification is effective.   

For the second parametrization the effects of the parameters on the matter power spectrum is fairly straight forward. As discussed above an increase (decrease) in $\mu$ will increase( decrease) the overall amplitude of the matter power spectrum. The parameter $\Sigma$ however only affects the low-$k$ matter power spectrum. A $\Sigma>1$($\Sigma<1$) will decrease(increase) the low-$k$ spectra relative to its $\Sigma=1$ value for a given $\mu$. This reflects the low-$k$ suppression(enhancement) of the growth for those parameter values. In the binned form of this parametrization the parameters within a given $k$-bin have the same effects as described above on the spectrum in that $k$-bin, and do not affect the other $k$-bin except near the transition between the bins. This allows parameters in different $z$-bins but the same $k$-bin to be degenerate with one another in their effect on the matter power spectrum today.
On the other hand, the effect of the first parametrization on the matter power spectrum is a bit more complicated due to the presence of the parameter $s$. As discussed above, just as $\mu$ increases or decreases the amplitude of the matter power spectrum in the second parametrization, the value of the parameter combination $Q R$ will increase or decrease the amplitude of the matter power spectrum. How much the amplitude is changed though is affected by the value of $s$. A larger $s$ means that the modification to gravity has not affected growth for very long, and thus the amplitude will be affected minimally. A smaller $s$ on the other hand means the modification has been present for a much longer time and thus the amplitude of the matter power spectrum will be affected most for a smaller $s$. A smaller $s$ will also affect the overall shape of the spectrum, mainly because when $s<1$ gets smaller it allows the modification to affect the evolution of the potential outside of the matter dominated era. For the low-$k$ spectrum as with the second parametrization, a $D>1$ (the parameter corresponding to $\Sigma$) will tend to cause a suppression in the low-$k$ spectrum  relative to its $D=1$ value (the converse still holds for $D<1$. A larger $s$, however, tends to prefer a suppression in the low-$k$ spectrum and can even overpower the boosting effect of $D<1$.  

Independent of parametrization changes to the matter power spectrum will of course affect the other observations such as the weak-lensing correlations, and the ISW-galaxy cross correlations. As discussed in \cite{Bean2010}, for example, values of MG parameters moderately higher than their GR values can lead to an anticorrelation in ISW-galaxy cross correlations at low redshift. Those values of MG parameters promote growth while suppressing the ISW effect. A more detailed review of the effect of different MG parameters on the ISW-effect in the CMB TT power spectrum is discussed in \cite{Bean2010,Daniel2010}.  Basically both smaller and \textit {sufficiently} larger values of MG parameters compared to their GR values will boost the ISW effect, while values of MG parameters \textit{ moderately} larger than their GR values will tend to suppress it. 
This is because the ISW effect is dependent on the square of
$\dot\phi+\dot\psi$.  A smaller (larger) value for the MG parameters will
increase (decrease) $\dot\phi+\dot\psi$ with larger values even causing the
quantity to go negative.  For sufficiently large MG parameters though
$\dot\phi+\dot\psi$ will become so negative that the ISW effect is again
boosted.

The tensions for both parametrizations are possibly due to the way the MG parameters affect the shape and amplitude of the matter power spectrum and the way this change affects the other observables.

In summary, our results using figures of merit and constraints from current data sets indicate that: 
\begin{enumerate}[(i)]
\item{Using the first parametrization with functional form, the strongest current constraints on the MG  parameters are found to be from the CMB+ISW+WL combination followed by the CMB+WL or CMB+MPK+ISW.}

\item{We observe, in the case of the first parametrization that the combinations CMB+WL or CMB+MPK do better than CMB+MPK+WL. Similarly, the combinations CMB+ISW+WL or CMB+MPK+ISW do better than the combination CMB+MPK+ISW+WL. A closer look at the best-fit models preferred by each combination of data sets shows that tensions appear between the best-fit values for the MG  parameters preferred by different data sets and that these tensions are much more pronounced when using this parametrization.}

\item{Using the second parametrization or the binning methods, we find that the combination CMB+MPK+ISW of current data sets consistently provides some of the strongest constraints on MG  parameters.}

\item{When using the second parametrization or the binning methods, we find that the combination CMB+MPK+ISW+WL provides, at best, only little improvement when compared to the combination CMB+MPK+ISW, again indicating some tension between the data sets. But we find here that the tension between the best-fit MG  parameters is much less pronounced than in the first parametrization although non-negligible.}

\item{In order to investigate these tensions, we fixed the core  cosmological parameters and left the MG  parameters to vary. In the cases of the second parametrization and the binning methods, the FoM's increase consistently (but only moderately in some cases) when adding a data set, indicating that some of the tension was reduced. A closer look at the best-fit MG  parameters preferred by different data sets shows that, while reduced, a non-negligible scatter is still present. This explains the only moderate improvements obtained in some cases. This also indicates that  there are non-negligible tensions intrinsic to the MG parameters preferred by different data sets. Moreover, using the first parametrization and fixing the core cosmological parameters did \emph{not} reduce the strong tensions observed between the MG  parameters which seems to suggest that the tensions are exacerbated by the functional form that the first parametrization is imposing on the MG  parameters.}

\item{We find that combining current data sets does not improve consistently the uncertainties on the MG  parameters due to tensions between the best-fit MG  parameters  {preferred} by different data sets. Some functional forms imposed by the parametrizations can lead to an exacerbation of these tensions and this point requires further future investigations.
It remains of interest to use these data sets separately in order to derive independent constraints on the MG parameters and thus allow one to cross-validate the results.}

\item{Unlike previous work that used a binned parametrization and the CFHTLS lensing data, we do not find a deviation from GR using a similar method but the refined HST-COSMOS data, confirming thus previous claims in those studies that the previous result were likely due to some systematic effect in using CFHTLS lensing data.}

\item{Finally, for all the parametrizations and binning methods used here, we find that the parameter values of general relativity are within the $95\%$ confidence level contours for all data set combinations.}
\end{enumerate}

\begin{center}
\begin{table}[t]
\begin{tabular}{|c|c|c|c|}\hline
\multicolumn{4}{|c|}{\bfseries Constraints for  $\{\mu_{i},\,\Sigma_{i}\}$ binned parametrization $0<z\leq1.5,\,1.5<z\leq3$ }\\ \hline
 &\multicolumn{3}{|c|}{$k_x=0.01$}\\ \hline
  data sets&WL &ISW&ISW, WL\\ \hline
  FoM$_{1}$ &  94.98 & 120.9 & 116.9 \\ \hline
$\qquad\mu_{1}\qquad$ & $\qquad[ 0.629,\,1.433]\qquad$ &$\qquad[ 0.537,\,1.197]\qquad$ & $\qquad[ 0.540,\,1.222]\qquad$\\ \hline
$\Sigma_{1}$ &  $[ 0.929,\,1.127]$ &$[ 0.948,\,1.137]$ & $[ 0.939,\,1.128]$\\ \hline 
$\eta_{1}$ &  $[ 0.4004,\,2.348]$ &$[ 0.702,\,2.992]$ & $[ 0.659,\,2.914]$\\
 (derived)&&&\\ \hline 
FoM$_{2}$ & 149.6 &165.7 & 163.5 \\ \hline   
$\mu_{2}$ & $[ 0.620,\,1.455]$ &$[ 0.570,\,1.352]$ & $[ 0.579,\,1.384]$ \\ \hline 
$\Sigma_{2}$ &  $[ 0.952,\,1.071]$ &$[ 0.959,\,1.074]$ & $[ 0.956,\,1.069]$ \\ \hline 
$\eta_{2}$ &  $[ 0.368,\,2.307]$ &$[ 0.485,\,2.616]$ & $[ 0.433,\,2.542]$\\
 (derived)&&&\\ \hline 
FoM$_{3}$ &  28.20 &40.89 & 46.09 \\ \hline 
$\mu_{3}$ & $[ 0.393,\,1.807]$ &$[ 0.594,\,1.446]$ &$[ 0.604,\,1.443]$ \\ \hline 
$\Sigma_{3}$ &   $[ 0.869,\,1.234]$ &$[ 0.795,\,1.209]$ & $[ 0.862,\,1.238]$ \\ \hline 
$\eta_{3}$ &  $[ 0.133,\,4.479]$ &$[ 0.288,\,2.489]$ & $[ 0.387,\,2.576]$\\ 
 (derived)&&&\\ \hline 
FoM$_{4}$ &  48.17 & 68.50 & 69.97 \\ \hline 
$\mu_{4}$ &  $[ 0.494,\,1.600]$ &$[ 0.545,\,1.270]$ & $[ 0.574,\,1.318]$ \\ \hline 
$\Sigma_{4}$ &  $[ 0.886,\,1.145]$ &$[ 0.858,\,1.133]$ & $[ 0.879,\,1.146]$  \\ \hline 
$\eta_{4}$ &  $[ 0.239,\,3.135]$ &$[ 0.509,\,2.687]$ & $[ 0.483,\,2.631]$\\ 
 (derived)&&&\\ \hline 
\hline

\hline
\end{tabular}
\caption{\label{table:obs4}
$95\%$ C.L. for the parameters $\mu_{i}$, $\eta_{i}$, and $\Sigma_{i}$ from  {the second parametrization \cite{GongBo2010,MGCAMB}} for redshift  and scale dependence as a binned $2 \times  2$ parametrization with $0<z\leq 1.5,\,1.5<z \leq 3$ and scale separation at $k_x= 0.01\, h\, Mpc^{-1}$.  The data used is described in the text above and indicated as: ISW-galaxy cross correlations (ISW), weak-lensing tomography (WL). In all cases, the data used is combined with WMAP7 temperature and polarization spectrum (CMB), the matter power spectrum (MPK), SN, BAO, BBN, AGE and $H_0$.  The FoM for each data set and $\{\mu_{i},\,\Sigma_{i}\}$ bin is also given.
}
\end{table}
\end{center}

\acknowledgements
{We thank T. Schrabback, G. B. Zhao, R. Bean, E. Linder, I. Tereno, D. Huterer, Y-S. Song, S. Daniel, and
P. Zhang for useful comments.  We are grateful to G. B. Zhao for making his code \texttt{MGCAMB} available to us. \texttt{MGCAMB} was modified and used for the second parametrization in this paper while we modified a full separate set of codes that we used for the first parametrization including a new modified version of \texttt{CAMB}, \texttt{CosmoMC}, as well as WL and ISW codes.  The codes that we developed are available from the authors at \cite{jnd}. M.I. acknowledges that this material is based upon work supported by Department of Energy (DOE) under grant DE-FG02-10ER41310 and that part of the calculations for this work have been performed on the Cosmology Computer Cluster funded by the Hoblitzelle Foundation.  J.D. acknowledges that this research was supported in part by the DOE Office of Science Graduate Fellowship Program (SCGF). The DOE SCGF Program was made possible in part by the American Recovery and Reinvestment Act of 2009.  The DOE SCGF program is administered by the Oak Ridge Institute for Science and Education for the DOE. ORISE is managed by Oak Ridge Associated Universities (ORAU) under DOE contract number DE-AC05-06OR23100.}


\begin{thebibliography}{}
\bibitem{CFHTNote} We intended in this work to apply the analysis to the CFHTLS weak-lensing data but it was recommended to us by leading members of the CFHTLS weak-lensing team to wait until the next data release because of known residual systematics in the T003 data release.
\end{thebibliography}
\end{document}